\documentclass[aps,prl,reprint,superscriptaddress]{revtex4-2}

\usepackage{float}
\usepackage{units}
\usepackage{graphicx}

\usepackage{upgreek}
\usepackage{xcolor}
\usepackage{graphicx}
\usepackage{amsmath}
\usepackage{amssymb}
\usepackage{cancel}
\usepackage{overpic}

\begin{document}

\title{Beyond Maxwell-Boltzmann statistics using confined vapor cells}

\author{Gilad Orr}
\affiliation{Department of Physics, Ariel University, Ariel 40700}
\author{Eliran Talker}
\email{elirant@ariel.ac.il}
\affiliation{Department of Electrical Engineering, Ariel University, Ariel 40700, Israel}

\begin{abstract}
Coherence time of thermal photons in rubidium vapor cells with varying
thicknesses, reveal that there is clear dependence of the photon correlation
time on cell thickness. Standard theoretical models accurately predict
the coherence time in centimeter-scale cells. In this study we demonstrated,
that these models break down in micrometer and sub-micrometer regimes.
Cell sizes ranging from mm-scale down to 200 nm did not adhere to
prediction based on the standard models. In order to address this
shortcoming, we develop an alternative approach better suited for
estimating photonic coherence times in ultra-thin vapor cells. This
work, highlights the need for a modified theoretical treatment of
the coherence time in the nanoscale regime.
\end{abstract}

\maketitle

\section{Introduction}

Understanding coherence and temporal dynamics of light is crucial as it governs light-light and light-matter interaction. Temporal coherence relates to the correlation of light with itself at different points in time, while spacial coherence describes the phase relationship between points in space during the light's wavefront propagation. This understanding is foundational to applied optics in many fields including holography, optical communications and medical imaging for example. It is regarded as first order correlation and is explained classically. As we have discussed the first order, a second order is implied, and the reader may be curious about the nature of the second order of the correlation function. The surprising answer is that it is quantum in nature at one of its extremes, and provides a fundamental insight to the nature of the light. This second order correlation, gives us the ability to distinguish light sources to classical, coherent and non-classical light sources. It is critical for characterizing quantum light sources \citep{glauber1963quantum,mandel1995optical}.  
Let us define the second-order intensity autocorrelation function of the optical
field, usually denoted by $g^{(2)}(\uptau)$,  as a measure of the correlation between the optical field intensity at time $t$ and at a later time $t+\uptau$. We essentially quantify the probability of detecting a photon after a time delay $\uptau$, given that a photon was detected at time $t$.
In particular, the value of this function at $\uptau=0$ reveals the nature of the light source:
\begin{description}
\item[$g^{(2)}(0)>1$] implies Gaussian statistics (photon bunching) a classical phenomena in which 'bunches' of photons are emitted from thermal light sources. This was first observed by Hanbury-Brown and Twiss experiment \citep{loudon2000quantum,brown1956correlation,brown1954lxxiv}, later applying it to determine the diameter of Sirius (star). It provides for calculating the coherence time and linewidth of thermal light \citep{arecchi1966measurement}.
\item[$g^{(2)}(0)=1$] implies Poisson statistics, as emitted from coherent sources such as lasers.
\item[ $g^{(2)}(0)<1$] implies non classical light (photon antibunching \citep{mandel1995optical}) which characterizes single photon sources. Observing antibunching demonstrated the quantum nature of light.
\end{description}    
Measurements of this function with fully resolved temporal dependence
became in the last decade a routine task in quantum information \citep{neergaard2006generation,neergaard2013quantum,singh2024quantum,takase2024generation,sychev2017enlargement}
and the characterization of a single-photon source such as quantum
dot \citep{arakawa2020progress}, single molecule \citep{toninelli2021single},
and diamond color center \citep{doherty2013nitrogen}, where its width
is in the nanosecond scale.

Recent advances in confining atoms and optical fields at micro and
nanoscale dimensions have opened new avenues for exploring fundamental
physics \citep{talker2019efficient,talker2017fluorescence,keaveney2012cooperative,delpy2025anomalous},
developing precision sensors \citep{talker2019efficient,carle2021exploring,liu2013ramsey},
and miniaturizing optical and quantum technologies. One intriguing
observation is that the coherence time of light scattered from hot
atomic vapor confined in submicron cells increases as the cell thickness
decreases. This behavior cannot be explained by conventional models
based on the Maxwell-Boltzmann velocity distribution alone. Instead,
it requires accounting for additional phenomena such as atomic quenching
due to wall collisions, and transit-time broadening.

In this work, we investigate the coherence time of thermal photons
in rubidium vapor cells of varying thicknesses, from centimeter-scale
down to 200 nm, by measuring the second-order intensity autocorrelation
function. Our results reveal a strong dependence of the photon correlation
time on cell thickness. While standard models accurately predict the
coherence time in centimeter-scale cells, they fail in the micrometer
and sub-micrometer regimes. To address this, we develop a refined
theoretical framework better suited for estimating photonic coherence
times in ultra-thin vapor cells, emphasizing the need for a revised
theoretical approach at the nanoscale. The paper is structured as
follows: we begin with a theoretical analysis using a Lindblad master
equation that incorporates various broadening and shifting mechanisms.
We then present our experimental results and discuss their implications.
Finally, we conclude with a summary of our findings.

\section{Theoretical Background}

The second order correlation function of light scattered from an illuminated
two-level atom is given by \citep{orszag2024quantum}
\begin{equation}
g^{(2)}(\uptau)=\frac{G^{(2)}(t,\uptau)}{\left|G^{(1)}(t)\right|^{2}}\label{eq:second_order_correlation_function}
\end{equation}

Where $G^{(1)}(\uptau)=\left\langle \hat{E}^{-}(t)\hat{E}^{+}(t)\right\rangle $,
and $G^{(2)}(t,\uptau)=\left\langle \hat{E}^{-}(t)\hat{E}^{-}(t+\uptau)\hat{E}^{+}(t)\hat{E}^{+}(t+\uptau)\right\rangle $
the the field operator has an infinite number of degrees of freedom,
hence, it might appear complicated to obtain an expression for $E^{+}(t)$.
However, we can take advantage of the existing result which is valid
for classically and quantum-mechanically that in the far field regime
the emitted field is proportional to the dipole approximation. Omitting
negligible constants and, for simplicity, ignoring the vectorial character
of the field in the operator we obtain a field that is proportional
to the dipole $d$ of the atom at time $t-\nicefrac{r}{c}$.
\[
E^{(+)}(r,t)\propto\frac{1}{r}(\rho_{eg})\left(t-\frac{r}{c}\right)
\]

Calculating $G^{(1)}(\uptau)$ and $G^{(2)}(t,\uptau)$ using the
above relation (Equation \ref{eq:second_order_correlation_function})
and quantum regression theorem \citep{khan2022quantum} we can write
the correlation function as
\begin{equation}
g^{(2)}(\uptau)=\frac{\rho_{22}^{ss}(\uptau)}{\rho_{22}^{ss}(\infty)}
\end{equation}

Where $\uptau$ is the time delay and $\rho_{22}^{ss}$ is the steady
state value of the excited state population. By assuming resonance
condition ($\Delta=0$) and low light intensity limit ($\Omega\rightarrow0$)
we can write the correlation function as (for a full derivation see
the Appendix)
\begin{equation}
g^{(2)}(\uptau)=1+\frac{\Gamma_{eff}^{2}\Gamma_{2}e^{-\Gamma_{eff}\uptau}-\Gamma_{eff}(2\Gamma_{2}-\Gamma_{eff})^{2}e^{-\Gamma_{2}\uptau}}{(\Gamma_{eff}-\Gamma_{2})(2\Gamma_{2}+\Gamma_{eff})^{2}}\label{eq:Derived_correlation_fuction}
\end{equation}

Where $\Gamma_{eff}$ is the effective linewidth and is equal to $\Gamma_{eff}^{2}\propto\left(\frac{\alpha}{t_{trans}}\right)^{2}$
with $\alpha$ a unitless factor and $t_{trans}$ is the transit time
broadening. $\Gamma_{2}$ is the emission rate which is associated
with the relaxation from the excited states to the ground states including
radiating and non-radiating process. The first decay element as shown
in Equation \ref{eq:Derived_correlation_fuction} is $\Gamma_{2}=\Gamma_{eff}/2+\Gamma_{dep}$
where $\Gamma_{dep}$ is the quenching rate from atom-wall collisions
which leads to nonradiative decay from the excited level to the ground
state level (see Supplemental Material \citep{supp}). Taking all these decay elements into
consideration one can see that the $g^{(2)}(\uptau)$ is highly dependent
on this parameter and can dramatically reduce the linewidth of the
measurements (for more information regarding the theory see Supplemental Material
part II \citep{supp}) 

\section{Experimental Methods}

\begin{figure}
\begin{centering}
\includegraphics[scale=0.5]{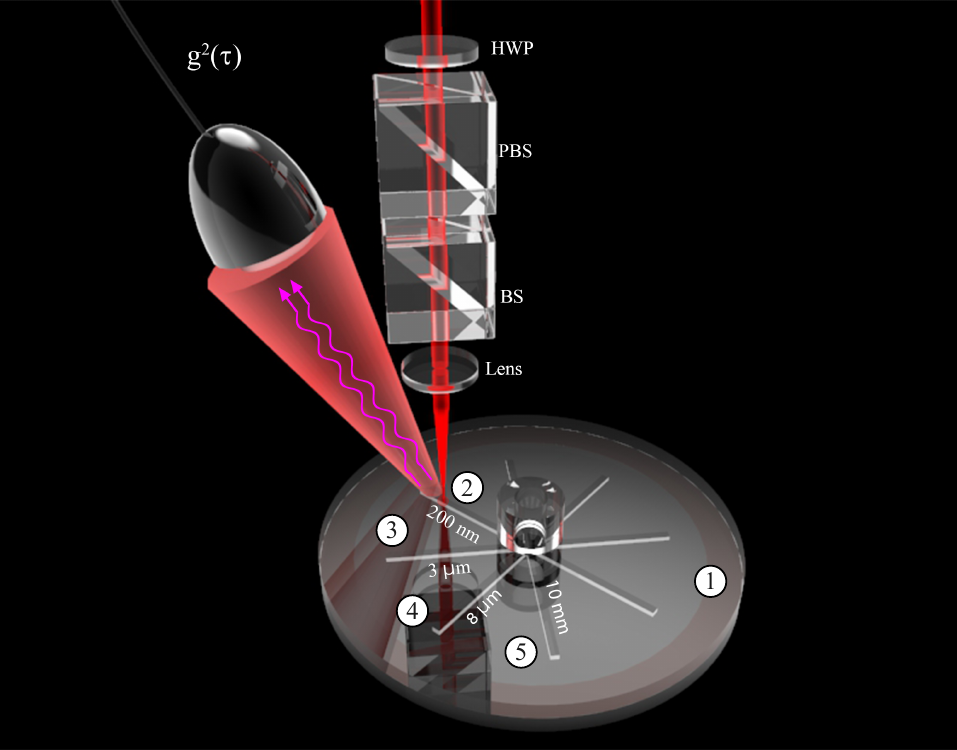}
\par\end{centering}
\caption{\label{fig:The-experimental-setup}3D illustration of the experimental
setup. The cell is made from 3 channels varying from $8\mu m$ to
$200\,nm$. The laser beam is stabilized to the close transition $F=3\rightarrow F'=4$
, laser beam power $50\mu W$ and the beam waist was set to $\sim7\mu m$.
\textbf{HWP} - half wave plate; \textbf{PBS} - polarizing beam splitter;
\textbf{BS} - non-polarizing beam splitter. A schematic layout of
our whole setup is presented in the Supplemental Material \citep{supp}.}

\end{figure}

The experimental setup is illustrated in Fig \ref{fig:The-experimental-setup}. 

At the heart of the setup is a vapor cell \textcircled{\tiny{1}}
containing three channels with a $10\,mm$ length and $1\,mm$ width,
with the thickness (height) distinguishing between the different channels.
In this experiment, the fabricated channel thicknesses were $200\,nm$,
$3\,\mu m,$and $8\,\mu m$ (\textcircled{\tiny{2}}, \textcircled{\tiny{3}}, and \textcircled{\tiny{4}}
in Fig. \ref{fig:The-experimental-setup} respectively). The cell
consists of two borosilicate (pyrex) disks. The first disk in which
the channels were fabricated using laser lithography, and etching.
Final surface treatment was obtained by reactive ion etching (RIE)
followed by deposition of polysilicon on the bonded surfaces. The
second disk seals the channels and accommodates the orifice required
for evacuating from air and filling of the cells with gas. In the
center of the second disk (cover disk), similar in size to the previous
one, a hole is drilled and a 4mm ID, glass tube was welded to the
cover disk acting as the orifice. The two disks were sealed to each
other using anodic bonding. The cell's channels were flushed and evacuated
after which they were filled with rubidium vapor and sealed by welding
the glass tube. A detailed account regarding the technique can be
found here \citep{talker2020atomic}). To this experiment we added
4 Rubidium cells supplied by \emph{Precision Glassblowing Colorado
USA} \citep{PrecisionGlassblowing} with cell thicknesses of $2\,mm,\,5\,mm,\,7.5\,mm$
and $10\,mm$. The cell is placed in a PID temperature controlled
oven \citep{orr2018high} specifically designed and fabricated by
us to avoid introducing any stray magnetic field. Such a magnetic
field results naturally from the current flowing through the heating
element. The heating element wire consisted of a twisted pair of fine
insulated wires with the current through each wire flowing in an opposite
direction. Fine micro-welding techniques were employed in fabricating
the furnace \citep{orr2012safe}. A 780nm, 100mW laser beam was used
as a probe. We used an external cavity diode (laser Toptica DL Pro)
passing the beam through a half wave plate followed by a polarizing
beam splitter. This allows us to control the beam intensity and improve
beam quality and polarization. The polarizer splits the beam by a
90:10 ratio, with the 10\% output beam propagating into a saturated
absorption spectroscopy (SAS) scheme \citep{debs2008piezo,wan2016laser,dammalapati2009saturated,liang2018laser}
to stabilize the laser frequency to the closed $(F=3\rightarrow F'=4)$
$D_{2}$ transition of rubidium 85, mimicking a two level system.
Even though we have observed Doppler broadening, our calculations
show that a two level system suggests a better agreement with the
experimental result. The second 90\% of the beam emerging from the
beam splitter, passes through a lens obtaining a beam waist of $100\mu m$
while projecting it on the different channels. The laser output power
was set to $50\mu W$. The scattered light is collected with a single
mode fiber (SMF), placed at a distance L=10 cm after the cell, at
an angle from the beam propagation direction. The scattered light
is focused on the SMF using lenses. The distance we set the fiber
is chosen such that the conditions for maximum spatial coherence are
satisfied \citep{nakayama2010precise}. Measuring the second-order
intensity autocorrelation is acheived by splitting the fiber coupled
light to a 50/50 ratio using a fiber beam splitter, feeding the two
beams (branches) to two single-photon avalanche photo diodes (Excelitas,
SPCM-800-12-FC). At $780nm$ the ADP's feature a quantum efficiency
of approximately 30\%. 

\section{Experimental results and discussion}

\begin{figure}
\begin{centering}
\includegraphics[scale=0.6]{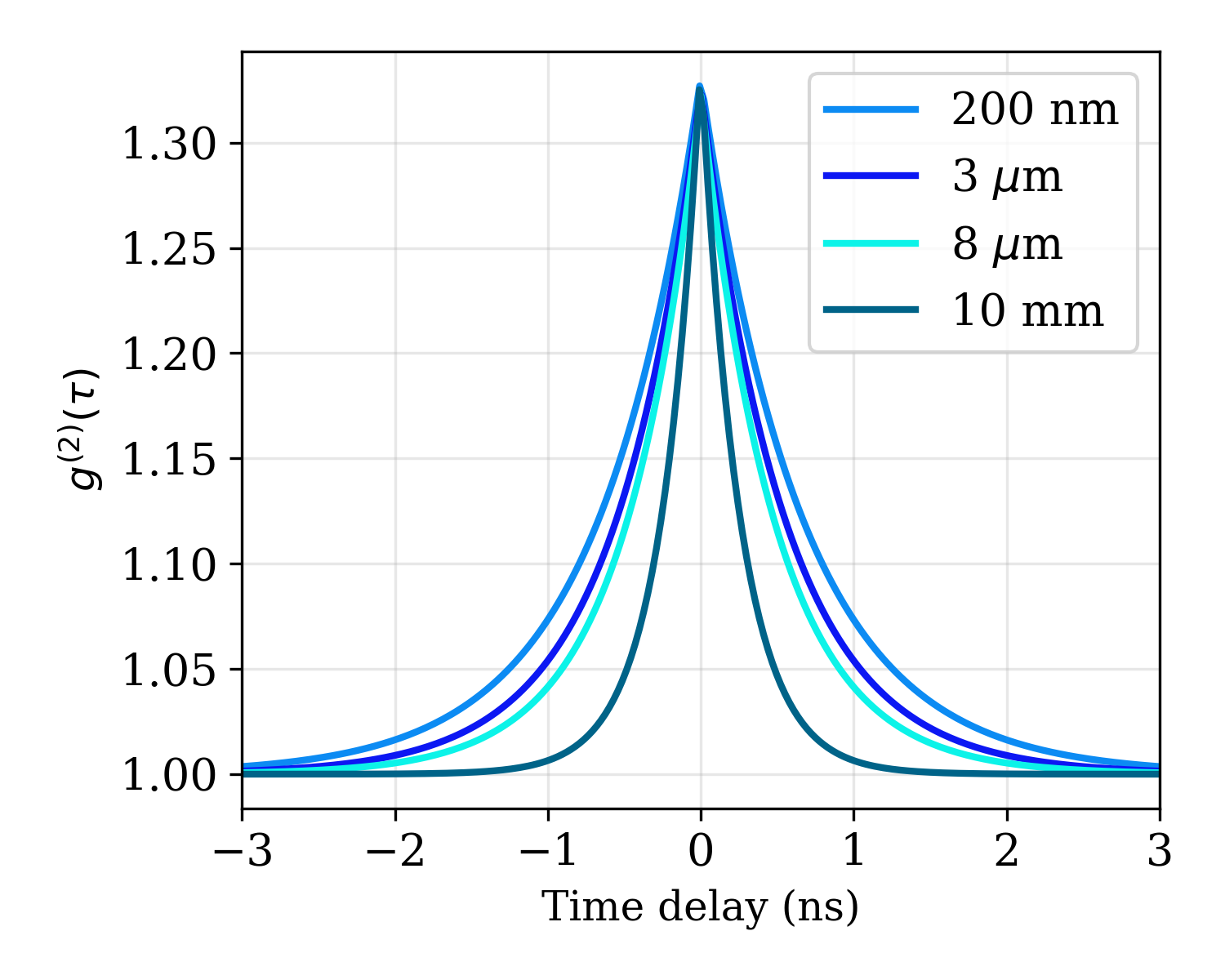}
\par\end{centering}
\caption{\label{fig:Theoretical-plot-of}Theoretical plot of the second-order
intensity autocorrelation as a function of the cell length, the whole
measurements done under low light intensity limit where the Rabi frequency
was near zero $(\Omega\rightarrow0)$ and the frequency detuning is
set to zero.}

\end{figure}

The theoretical plot of the second-order autocorrelation intensity
as a function of the cell thickness is seen in figure \ref{fig:Theoretical-plot-of}.
We observed that by decreasing the cell thickness the linewidth of
the $g^{(2)}(\uptau)$ measurements are increasing. 

The two main factors contributing to the change in the line width,
are the mean atomic velocity and the mean free path which are influence
by the cell thickness and attributed to the dephasing rate given by
\citep{talker2020atomic}
\begin{equation}
\Gamma_{dephase}=\frac{\bar{v}}{\bar{l}}+\frac{v_{z}}{L/2}
\end{equation}

Where $\bar{v}=\sqrt{\frac{2k_{B}T}{m_{Rb}}}$ is the mean atomic
velocity, $\bar{l}$ is the mean free path, $v_{z}$ is the atomic
velocity in the direction perpendicular to the cell walls, and L is
the cell thickness. Another, recent approach for calculating the velocities in gases is presented by G{\'a}bor \citep{lente2025direction}.  The mean free path $\bar{l}=\nicefrac{\bar{v}_{rel}}{\Gamma_{se}}$,
where $\bar{v}_{rel}$ is the relative mean thermal velocity if the
rubidium atoms, and the rate of spin exchange $\Gamma_{se}=\sigma_{se}\cdot\bar{v}_{rel}\cdot n_{Rb}$,
$\sigma_{se}=2\times10^{-14}cm^{2}$ is the spin exchange cross section,
$\bar{v}_{rel}=\left(\frac{8k_{b}T}{\pi m_{Rb}}\right)$ is the average
relative velocity of the rubidum atoms $m_{Rb}$ is the reduced mass
of the system of two rubidium atoms, and $n_{Rb}$ is the density
of atoms in the cell (for more information see Supplemental Material \citep{supp}). 

To calculate the unitless factor $\alpha$ which the effective linewidth
$\Gamma_{eff}$ in Equation \ref{eq:Derived_correlation_fuction}
relates to, we performed a Monte Carlo simulation of thermal atoms
traveling within the detection volume. The simulated results are plotted
with the fluorescence spectra obtain from different cell thickness
(see Supplemental Material Part I \citep{supp}). 

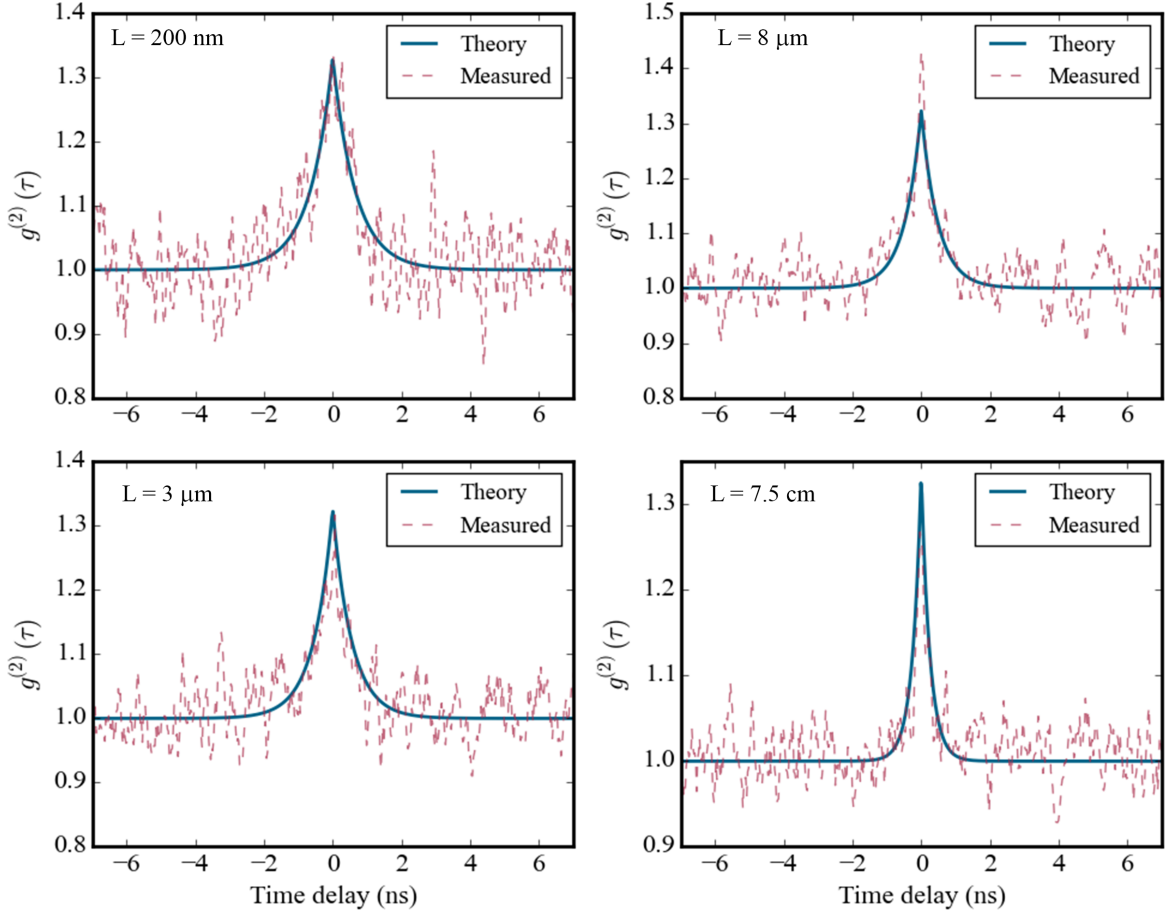
\begin{figure*}
\begin{centering}
 \begin{overpic}[width=0.8\textwidth]{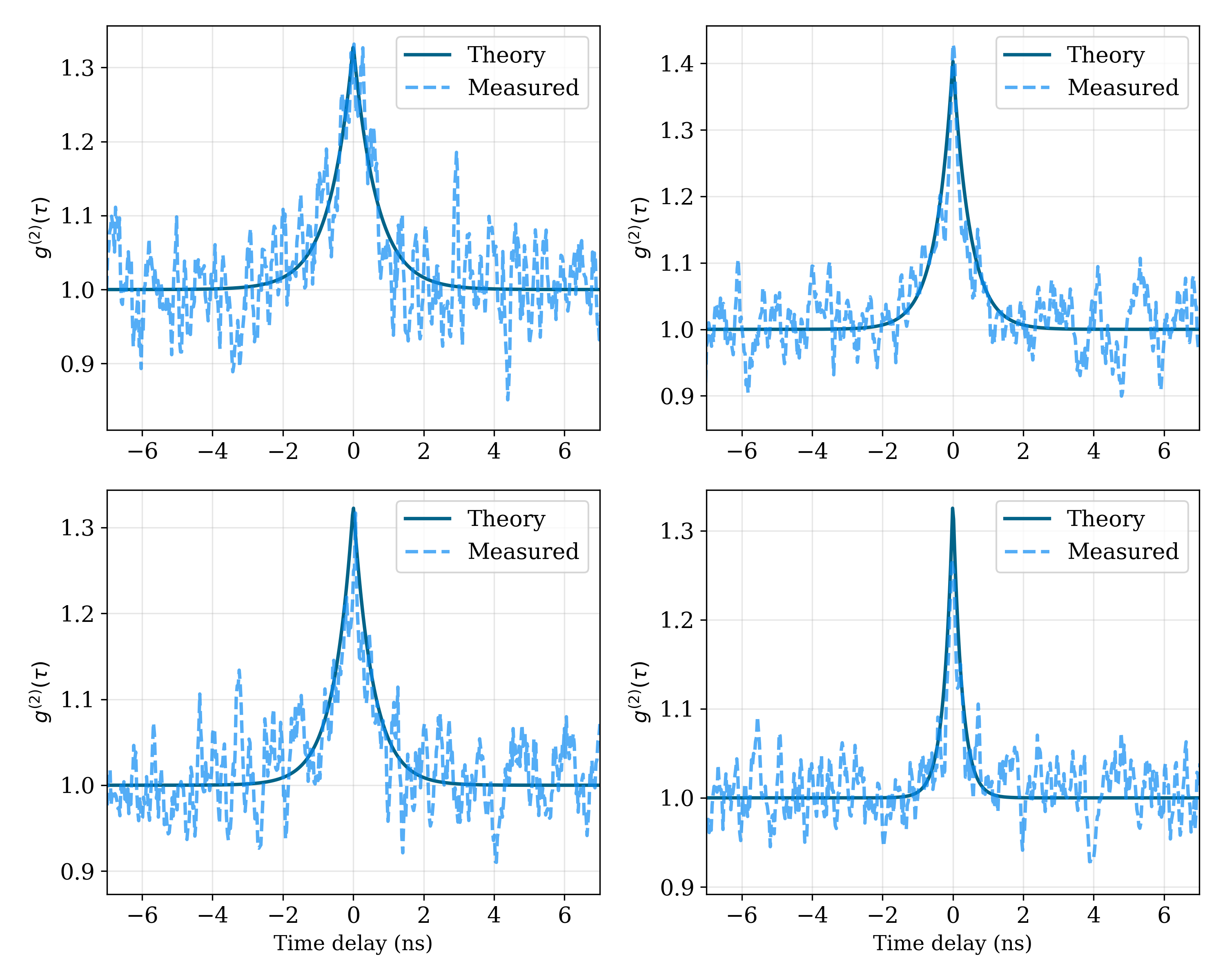}
        \put(10,75){\textcircled{\footnotesize{a}} $L=200nm$}
        \put(10,37){\textcircled{\footnotesize{b}} $L=3\mu m$}
        \put(60,75){\textcircled{\footnotesize{c}} $L=8\mu m$}
        \put(60,37){\textcircled{\footnotesize{d}} $L=75mm$}
 \end{overpic}
\par\end{centering}
\caption{\label{fig:Measurements-of-the-second-order-intensity-autocorrelation}Measurements
of the second-order intensity autocorrelation for different cell thickness
(a) $200nm$ (b) $3\mu m$ (c) $8\mu m$ (d) 75 mm. The calculated
result for the different cell thicknesses, based on equation \ref{eq:Derived_correlation_fuction}
is shown to be in excellent agreement.}

\end{figure*}

Comparing the theoretical results obtained from Equation \ref{eq:Derived_correlation_fuction}
to the experimental data (Figure \ref{fig:Measurements-of-the-second-order-intensity-autocorrelation}),
we find excellent agreement with the measurements obtained from different
cell thicknesses. The theoretical plotted result was obtained by inserting
all the required decay factors into Equation \ref{eq:Derived_correlation_fuction}
. While for mm size cell thicknesses some of the decay factors may
be ignored, for cell thicknesses below $50\mu m$ they are required. 

The transit time broadening calculation was based on a set temperature
of $150^{o}C$ resulting in an atomic density of approximately $3.5\times10^{19}cm^{-3}$
, and on a mean speed of $280\,m/s$.

We find that our simple simulation is sufficient to capture the experimentally
observed trend with the unitless factor $\alpha=8.25$. 

We will now turn to calculating the coherence length of the thermal
photons
\begin{equation}
L_{co}=c\cdot\uptau_{co}
\end{equation}

\begin{figure*}
\centering{}\includegraphics[width=0.5\textwidth]{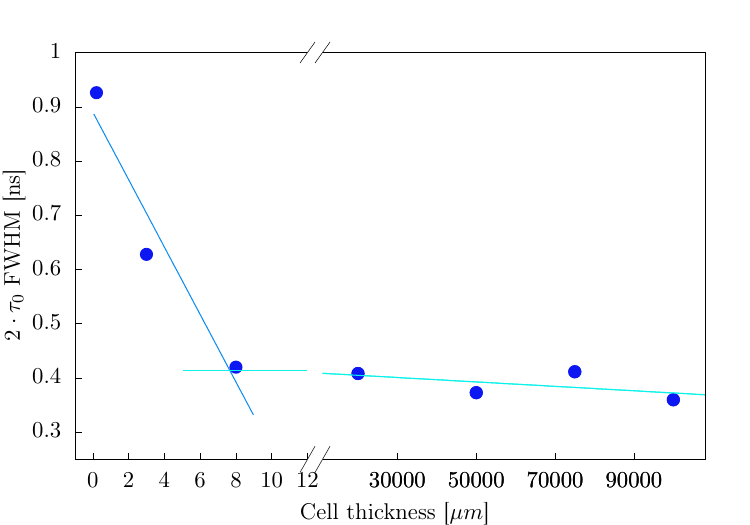}\includegraphics[width=0.5\textwidth]{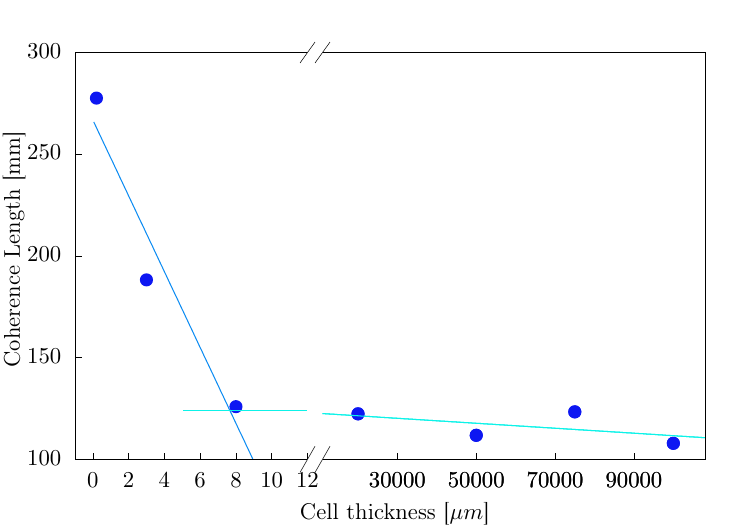}\caption{\label{fig:Measured-coherence}Measured coherence (a) time, as a function
of cell length (b) length, as a function of the cell length.}
\end{figure*}

Where c is the speed of light and $\uptau_{co}$ is the coherence
time from $g^{(2)}(\uptau)$ measurements obtained from both figure
\ref{fig:Theoretical-plot-of} and figure \ref{fig:Measurements-of-the-second-order-intensity-autocorrelation}.
As demonstrated, we can calculate the coherence length of the thermal
light which is scattered from hot rubidium vapor. Simulation based
on equation \ref{eq:Derived_correlation_fuction} shows that varying
the cell thickness from $100\mu m$ up to $20mm$ the coherence time
does not change dramatically with the observed average coherence time
being around $0.442\,ns$ (see Supplemental Material Equation
32 \citep{supp}). This changes abruptly, when the cell thickness goes below $8\mu m$
and reaching $3\mu m$ we start to observe that the average coherence
time is doubled (See figure \ref{fig:Measured-coherence}(a)) which
implies that we start seeing a limit to the thermal state coherence
time as obtained from the Maxwell-Boltzmann derivation. 

This result corresponds to the fact that if we compare $\Gamma_{eff}$
to $\Gamma_{dephase}$, $\Gamma_{eff}\approx2\pi\frac{\alpha}{t_{trans}}$
and $\Gamma_{dephase}\approx\frac{2v_{z}}{L}$, while taking into
account that $t_{trans}=\frac{D}{\bar{v}}$ (or $\Gamma_{eff}\approx\frac{\alpha\bar{v}}{D}$)
we can see that the coherence time of the system depends on the cell
thickness compared to the beam waist diameter. As long as $D<L$,
$\Gamma_{eff}$ is dominant and the coherent time is relatively constant.
But as $L$ decreases below the beam waist diameter $D$ the coherence
time increases. For similar diameters they produced similar coherence
times. The best result that we obtained with our rubidium cell is
coherence length of around 280 nm. This can improved if we continue
to reduce the cell length and increase the cell temperature leading
to an increase in the wall collision rate. This results in an increase
of the coherence time. Coherence length time and length increase come
at a price. The signal to noise ratio decreases since there is a decrease
in light matter interaction, and hence intensity, for the dominant
process will be a non-radiative decay due to quenching effect of the
cell walls.

\section{Conclusions }

We investigated the coherence time of thermal photons in rubidium
vapor cells. This work focused on cells with thicknesses ranging from
centimeter scale down to 200 nm, revealing a significant dependence
on cell thickness. Our measurements of the second-order intensity
autocorrelation function, ( $g^{(2)}(\uptau)$ ), demonstrated that
coherence time increases as cell thickness decreases, particularly
in the micrometer and submicrometer regimes. While standard theoretical
models based on Maxwell-Boltzmann velocity distributions accurately
predict coherence times in larger cells, they fail to account for
the observed behavior in cell thicknesses below $8\mu m$. This discrepancy
necessitated a refined theoretical framework incorporating transit-time
broadening and atomic quenching due to wall collisions, as described
by our Lindblad master equation approach. Our experimental results,
showed excellent agreement with the proposed model, capturing the
impact of cell thickness on linewidth and coherence properties. Notably,
for cells below $8\mu m$, coherence time increased by a factor of
two, with the smallest 200 nm cell achieving a coherence length of
approximately 290 nm. These findings highlight the limit of the Maxwell-Boltzmann
based model of thermal state coherence in confined systems. Also,
it underscores the SNR limit of non-radiative decay processes , such
as quenching, in nanoscale environments. This work opens new avenues
for understanding photon statistics in confined atomic systems, with
implications for developing miniaturized quantum technologies, precision
sensors, and optical devices. Future research could explore optimizing
cell design, by further reducing thickness or increasing cell temperature,
to enhance coherence times. This may come at a price of reduced signal-to-noise
ratios due to increased non-radiative decay. These insights provide
a foundation for advancing nanoscale photonics and quantum optics
applications. 

\bibliographystyle{apsrev4-2}
\bibliography{References}

\begin{thebibliography}{35}%
\makeatletter
\providecommand \@ifxundefined [1]{%
 \@ifx{#1\undefined}
}%
\providecommand \@ifnum [1]{%
 \ifnum #1\expandafter \@firstoftwo
 \else \expandafter \@secondoftwo
 \fi
}%
\providecommand \@ifx [1]{%
 \ifx #1\expandafter \@firstoftwo
 \else \expandafter \@secondoftwo
 \fi
}%
\providecommand \natexlab [1]{#1}%
\providecommand \enquote  [1]{``#1''}%
\providecommand \bibnamefont  [1]{#1}%
\providecommand \bibfnamefont [1]{#1}%
\providecommand \citenamefont [1]{#1}%
\providecommand \href@noop [0]{\@secondoftwo}%
\providecommand \href [0]{\begingroup \@sanitize@url \@href}%
\providecommand \@href[1]{\@@startlink{#1}\@@href}%
\providecommand \@@href[1]{\endgroup#1\@@endlink}%
\providecommand \@sanitize@url [0]{\catcode `\\12\catcode `\$12\catcode
  `\&12\catcode `\#12\catcode `\^12\catcode `\_12\catcode `\%12\relax}%
\providecommand \@@startlink[1]{}%
\providecommand \@@endlink[0]{}%
\providecommand \url  [0]{\begingroup\@sanitize@url \@url }%
\providecommand \@url [1]{\endgroup\@href {#1}{\urlprefix }}%
\providecommand \urlprefix  [0]{URL }%
\providecommand \Eprint [0]{\href }%
\providecommand \doibase [0]{https://doi.org/}%
\providecommand \selectlanguage [0]{\@gobble}%
\providecommand \bibinfo  [0]{\@secondoftwo}%
\providecommand \bibfield  [0]{\@secondoftwo}%
\providecommand \translation [1]{[#1]}%
\providecommand \BibitemOpen [0]{}%
\providecommand \bibitemStop [0]{}%
\providecommand \bibitemNoStop [0]{.\EOS\space}%
\providecommand \EOS [0]{\spacefactor3000\relax}%
\providecommand \BibitemShut  [1]{\csname bibitem#1\endcsname}%
\let\auto@bib@innerbib\@empty
\bibitem [{\citenamefont {Glauber}(1963)}]{glauber1963quantum}%
  \BibitemOpen
  \bibfield  {author} {\bibinfo {author} {\bibfnamefont {R.~J.}\ \bibnamefont
  {Glauber}},\ }\href@noop {} {\bibfield  {journal} {\bibinfo  {journal}
  {Physical Review}\ }\textbf {\bibinfo {volume} {130}},\ \bibinfo {pages}
  {2529} (\bibinfo {year} {1963})}\BibitemShut {NoStop}%
\bibitem [{\citenamefont {Mandel}\ and\ \citenamefont
  {Wolf}(1995)}]{mandel1995optical}%
  \BibitemOpen
  \bibfield  {author} {\bibinfo {author} {\bibfnamefont {L.}~\bibnamefont
  {Mandel}}\ and\ \bibinfo {author} {\bibfnamefont {E.}~\bibnamefont {Wolf}},\
  }\href@noop {} {\emph {\bibinfo {title} {Optical coherence and quantum
  optics}}}\ (\bibinfo  {publisher} {Cambridge university press},\ \bibinfo
  {year} {1995})\BibitemShut {NoStop}%
\bibitem [{\citenamefont {Loudon}(2000)}]{loudon2000quantum}%
  \BibitemOpen
  \bibfield  {author} {\bibinfo {author} {\bibfnamefont {R.}~\bibnamefont
  {Loudon}},\ }\href@noop {} {\emph {\bibinfo {title} {The quantum theory of
  light}}}\ (\bibinfo  {publisher} {OUP Oxford},\ \bibinfo {year}
  {2000})\BibitemShut {NoStop}%
\bibitem [{\citenamefont {Brown}\ and\ \citenamefont
  {Twiss}(1956)}]{brown1956correlation}%
  \BibitemOpen
  \bibfield  {author} {\bibinfo {author} {\bibfnamefont {R.~H.}\ \bibnamefont
  {Brown}}\ and\ \bibinfo {author} {\bibfnamefont {R.~Q.}\ \bibnamefont
  {Twiss}},\ }\href@noop {} {\bibfield  {journal} {\bibinfo  {journal}
  {Nature}\ }\textbf {\bibinfo {volume} {177}},\ \bibinfo {pages} {27}
  (\bibinfo {year} {1956})}\BibitemShut {NoStop}%
\bibitem [{\citenamefont {Brown}\ and\ \citenamefont
  {Twiss}(1954)}]{brown1954lxxiv}%
  \BibitemOpen
  \bibfield  {author} {\bibinfo {author} {\bibfnamefont {R.~H.}\ \bibnamefont
  {Brown}}\ and\ \bibinfo {author} {\bibfnamefont {R.~Q.}\ \bibnamefont
  {Twiss}},\ }\href@noop {} {\bibfield  {journal} {\bibinfo  {journal} {The
  London, Edinburgh, and Dublin Philosophical Magazine and Journal of Science}\
  }\textbf {\bibinfo {volume} {45}},\ \bibinfo {pages} {663} (\bibinfo {year}
  {1954})}\BibitemShut {NoStop}%
\bibitem [{\citenamefont {Arecchi}\ \emph {et~al.}(1966)\citenamefont
  {Arecchi}, \citenamefont {Berne},\ and\ \citenamefont
  {Sona}}]{arecchi1966measurement}%
  \BibitemOpen
  \bibfield  {author} {\bibinfo {author} {\bibfnamefont {F.}~\bibnamefont
  {Arecchi}}, \bibinfo {author} {\bibfnamefont {A.}~\bibnamefont {Berne}},\
  and\ \bibinfo {author} {\bibfnamefont {A.}~\bibnamefont {Sona}},\ }\href@noop
  {} {\bibfield  {journal} {\bibinfo  {journal} {Physical Review Letters}\
  }\textbf {\bibinfo {volume} {17}},\ \bibinfo {pages} {260} (\bibinfo {year}
  {1966})}\BibitemShut {NoStop}%
\bibitem [{\citenamefont {Neergaard-Nielsen}\ \emph {et~al.}(2006)\citenamefont
  {Neergaard-Nielsen}, \citenamefont {Nielsen}, \citenamefont {Hettich},
  \citenamefont {M{\o}lmer},\ and\ \citenamefont
  {Polzik}}]{neergaard2006generation}%
  \BibitemOpen
  \bibfield  {author} {\bibinfo {author} {\bibfnamefont {J.~S.}\ \bibnamefont
  {Neergaard-Nielsen}}, \bibinfo {author} {\bibfnamefont {B.~M.}\ \bibnamefont
  {Nielsen}}, \bibinfo {author} {\bibfnamefont {C.}~\bibnamefont {Hettich}},
  \bibinfo {author} {\bibfnamefont {K.}~\bibnamefont {M{\o}lmer}},\ and\
  \bibinfo {author} {\bibfnamefont {E.~S.}\ \bibnamefont {Polzik}},\
  }\href@noop {} {\bibfield  {journal} {\bibinfo  {journal} {Physical review
  letters}\ }\textbf {\bibinfo {volume} {97}},\ \bibinfo {pages} {083604}
  (\bibinfo {year} {2006})}\BibitemShut {NoStop}%
\bibitem [{\citenamefont {Neergaard-Nielsen}\ \emph {et~al.}(2013)\citenamefont
  {Neergaard-Nielsen}, \citenamefont {Eto}, \citenamefont {Lee}, \citenamefont
  {Jeong},\ and\ \citenamefont {Sasaki}}]{neergaard2013quantum}%
  \BibitemOpen
  \bibfield  {author} {\bibinfo {author} {\bibfnamefont {J.~S.}\ \bibnamefont
  {Neergaard-Nielsen}}, \bibinfo {author} {\bibfnamefont {Y.}~\bibnamefont
  {Eto}}, \bibinfo {author} {\bibfnamefont {C.-W.}\ \bibnamefont {Lee}},
  \bibinfo {author} {\bibfnamefont {H.}~\bibnamefont {Jeong}},\ and\ \bibinfo
  {author} {\bibfnamefont {M.}~\bibnamefont {Sasaki}},\ }\href@noop {}
  {\bibfield  {journal} {\bibinfo  {journal} {Nature Photonics}\ }\textbf
  {\bibinfo {volume} {7}},\ \bibinfo {pages} {439} (\bibinfo {year}
  {2013})}\BibitemShut {NoStop}%
\bibitem [{\citenamefont {Singh}\ and\ \citenamefont
  {Teretenkov}(2024)}]{singh2024quantum}%
  \BibitemOpen
  \bibfield  {author} {\bibinfo {author} {\bibfnamefont {R.}~\bibnamefont
  {Singh}}\ and\ \bibinfo {author} {\bibfnamefont {A.~E.}\ \bibnamefont
  {Teretenkov}},\ }\href@noop {} {\bibfield  {journal} {\bibinfo  {journal}
  {Physics Open}\ }\textbf {\bibinfo {volume} {18}},\ \bibinfo {pages} {100198}
  (\bibinfo {year} {2024})}\BibitemShut {NoStop}%
\bibitem [{\citenamefont {Takase}\ \emph {et~al.}(2024)\citenamefont {Takase},
  \citenamefont {Hanamura}, \citenamefont {Nagayoshi}, \citenamefont
  {Bourassa}, \citenamefont {Alexander}, \citenamefont {Kawasaki},
  \citenamefont {Asavanant}, \citenamefont {Endo},\ and\ \citenamefont
  {Furusawa}}]{takase2024generation}%
  \BibitemOpen
  \bibfield  {author} {\bibinfo {author} {\bibfnamefont {K.}~\bibnamefont
  {Takase}}, \bibinfo {author} {\bibfnamefont {F.}~\bibnamefont {Hanamura}},
  \bibinfo {author} {\bibfnamefont {H.}~\bibnamefont {Nagayoshi}}, \bibinfo
  {author} {\bibfnamefont {J.~E.}\ \bibnamefont {Bourassa}}, \bibinfo {author}
  {\bibfnamefont {R.~N.}\ \bibnamefont {Alexander}}, \bibinfo {author}
  {\bibfnamefont {A.}~\bibnamefont {Kawasaki}}, \bibinfo {author}
  {\bibfnamefont {W.}~\bibnamefont {Asavanant}}, \bibinfo {author}
  {\bibfnamefont {M.}~\bibnamefont {Endo}},\ and\ \bibinfo {author}
  {\bibfnamefont {A.}~\bibnamefont {Furusawa}},\ }\href@noop {} {\bibfield
  {journal} {\bibinfo  {journal} {Physical Review A}\ }\textbf {\bibinfo
  {volume} {110}},\ \bibinfo {pages} {012436} (\bibinfo {year}
  {2024})}\BibitemShut {NoStop}%
\bibitem [{\citenamefont {Sychev}\ \emph {et~al.}(2017)\citenamefont {Sychev},
  \citenamefont {Ulanov}, \citenamefont {Pushkina}, \citenamefont {Richards},
  \citenamefont {Fedorov},\ and\ \citenamefont
  {Lvovsky}}]{sychev2017enlargement}%
  \BibitemOpen
  \bibfield  {author} {\bibinfo {author} {\bibfnamefont {D.~V.}\ \bibnamefont
  {Sychev}}, \bibinfo {author} {\bibfnamefont {A.~E.}\ \bibnamefont {Ulanov}},
  \bibinfo {author} {\bibfnamefont {A.~A.}\ \bibnamefont {Pushkina}}, \bibinfo
  {author} {\bibfnamefont {M.~W.}\ \bibnamefont {Richards}}, \bibinfo {author}
  {\bibfnamefont {I.~A.}\ \bibnamefont {Fedorov}},\ and\ \bibinfo {author}
  {\bibfnamefont {A.~I.}\ \bibnamefont {Lvovsky}},\ }\href@noop {} {\bibfield
  {journal} {\bibinfo  {journal} {Nature Photonics}\ }\textbf {\bibinfo
  {volume} {11}},\ \bibinfo {pages} {379} (\bibinfo {year} {2017})}\BibitemShut
  {NoStop}%
\bibitem [{\citenamefont {Arakawa}\ and\ \citenamefont
  {Holmes}(2020)}]{arakawa2020progress}%
  \BibitemOpen
  \bibfield  {author} {\bibinfo {author} {\bibfnamefont {Y.}~\bibnamefont
  {Arakawa}}\ and\ \bibinfo {author} {\bibfnamefont {M.~J.}\ \bibnamefont
  {Holmes}},\ }\href@noop {} {\bibfield  {journal} {\bibinfo  {journal}
  {Applied Physics Reviews}\ }\textbf {\bibinfo {volume} {7}} (\bibinfo {year}
  {2020})}\BibitemShut {NoStop}%
\bibitem [{\citenamefont {Toninelli}\ \emph {et~al.}(2021)\citenamefont
  {Toninelli}, \citenamefont {Gerhardt}, \citenamefont {Clark}, \citenamefont
  {Reserbat-Plantey}, \citenamefont {G{\"o}tzinger}, \citenamefont
  {Ristanovi{\'c}}, \citenamefont {Colautti}, \citenamefont {Lombardi},
  \citenamefont {Major}, \citenamefont {Deperasi{\'n}ska} \emph
  {et~al.}}]{toninelli2021single}%
  \BibitemOpen
  \bibfield  {author} {\bibinfo {author} {\bibfnamefont {C.}~\bibnamefont
  {Toninelli}}, \bibinfo {author} {\bibfnamefont {I.}~\bibnamefont {Gerhardt}},
  \bibinfo {author} {\bibfnamefont {A.}~\bibnamefont {Clark}}, \bibinfo
  {author} {\bibfnamefont {A.}~\bibnamefont {Reserbat-Plantey}}, \bibinfo
  {author} {\bibfnamefont {S.}~\bibnamefont {G{\"o}tzinger}}, \bibinfo {author}
  {\bibfnamefont {Z.}~\bibnamefont {Ristanovi{\'c}}}, \bibinfo {author}
  {\bibfnamefont {M.}~\bibnamefont {Colautti}}, \bibinfo {author}
  {\bibfnamefont {P.}~\bibnamefont {Lombardi}}, \bibinfo {author}
  {\bibfnamefont {K.}~\bibnamefont {Major}}, \bibinfo {author} {\bibfnamefont
  {I.}~\bibnamefont {Deperasi{\'n}ska}}, \emph {et~al.},\ }\href@noop {}
  {\bibfield  {journal} {\bibinfo  {journal} {Nature Materials}\ }\textbf
  {\bibinfo {volume} {20}},\ \bibinfo {pages} {1615} (\bibinfo {year}
  {2021})}\BibitemShut {NoStop}%
\bibitem [{\citenamefont {Doherty}\ \emph {et~al.}(2013)\citenamefont
  {Doherty}, \citenamefont {Manson}, \citenamefont {Delaney}, \citenamefont
  {Jelezko}, \citenamefont {Wrachtrup},\ and\ \citenamefont
  {Hollenberg}}]{doherty2013nitrogen}%
  \BibitemOpen
  \bibfield  {author} {\bibinfo {author} {\bibfnamefont {M.~W.}\ \bibnamefont
  {Doherty}}, \bibinfo {author} {\bibfnamefont {N.~B.}\ \bibnamefont {Manson}},
  \bibinfo {author} {\bibfnamefont {P.}~\bibnamefont {Delaney}}, \bibinfo
  {author} {\bibfnamefont {F.}~\bibnamefont {Jelezko}}, \bibinfo {author}
  {\bibfnamefont {J.}~\bibnamefont {Wrachtrup}},\ and\ \bibinfo {author}
  {\bibfnamefont {L.~C.}\ \bibnamefont {Hollenberg}},\ }\href@noop {}
  {\bibfield  {journal} {\bibinfo  {journal} {Physics Reports}\ }\textbf
  {\bibinfo {volume} {528}},\ \bibinfo {pages} {1} (\bibinfo {year}
  {2013})}\BibitemShut {NoStop}%
\bibitem [{\citenamefont {Talker}\ \emph {et~al.}(2019)\citenamefont {Talker},
  \citenamefont {Arora}, \citenamefont {Barash}, \citenamefont {Wilkowski},\
  and\ \citenamefont {Levy}}]{talker2019efficient}%
  \BibitemOpen
  \bibfield  {author} {\bibinfo {author} {\bibfnamefont {E.}~\bibnamefont
  {Talker}}, \bibinfo {author} {\bibfnamefont {P.}~\bibnamefont {Arora}},
  \bibinfo {author} {\bibfnamefont {Y.}~\bibnamefont {Barash}}, \bibinfo
  {author} {\bibfnamefont {D.}~\bibnamefont {Wilkowski}},\ and\ \bibinfo
  {author} {\bibfnamefont {U.}~\bibnamefont {Levy}},\ }\href@noop {} {\bibfield
   {journal} {\bibinfo  {journal} {Optics Express}\ }\textbf {\bibinfo {volume}
  {27}},\ \bibinfo {pages} {33445} (\bibinfo {year} {2019})}\BibitemShut
  {NoStop}%
\bibitem [{\citenamefont {Talker}\ \emph {et~al.}(2017)\citenamefont {Talker},
  \citenamefont {Stern}, \citenamefont {Naiman}, \citenamefont {Barash},\ and\
  \citenamefont {Levy}}]{talker2017fluorescence}%
  \BibitemOpen
  \bibfield  {author} {\bibinfo {author} {\bibfnamefont {E.}~\bibnamefont
  {Talker}}, \bibinfo {author} {\bibfnamefont {L.}~\bibnamefont {Stern}},
  \bibinfo {author} {\bibfnamefont {A.}~\bibnamefont {Naiman}}, \bibinfo
  {author} {\bibfnamefont {Y.}~\bibnamefont {Barash}},\ and\ \bibinfo {author}
  {\bibfnamefont {U.}~\bibnamefont {Levy}},\ }\href@noop {} {\bibfield
  {journal} {\bibinfo  {journal} {Journal of Physics Communications}\ }\textbf
  {\bibinfo {volume} {1}},\ \bibinfo {pages} {055016} (\bibinfo {year}
  {2017})}\BibitemShut {NoStop}%
\bibitem [{\citenamefont {Keaveney}\ \emph {et~al.}(2012)\citenamefont
  {Keaveney}, \citenamefont {Sargsyan}, \citenamefont {Krohn}, \citenamefont
  {Hughes}, \citenamefont {Sarkisyan},\ and\ \citenamefont
  {Adams}}]{keaveney2012cooperative}%
  \BibitemOpen
  \bibfield  {author} {\bibinfo {author} {\bibfnamefont {J.}~\bibnamefont
  {Keaveney}}, \bibinfo {author} {\bibfnamefont {A.}~\bibnamefont {Sargsyan}},
  \bibinfo {author} {\bibfnamefont {U.}~\bibnamefont {Krohn}}, \bibinfo
  {author} {\bibfnamefont {I.~G.}\ \bibnamefont {Hughes}}, \bibinfo {author}
  {\bibfnamefont {D.}~\bibnamefont {Sarkisyan}},\ and\ \bibinfo {author}
  {\bibfnamefont {C.~S.}\ \bibnamefont {Adams}},\ }\href@noop {} {\bibfield
  {journal} {\bibinfo  {journal} {Physical review letters}\ }\textbf {\bibinfo
  {volume} {108}},\ \bibinfo {pages} {173601} (\bibinfo {year}
  {2012})}\BibitemShut {NoStop}%
\bibitem [{\citenamefont {Delpy}\ \emph {et~al.}(2025)\citenamefont {Delpy},
  \citenamefont {Fayard}, \citenamefont {Bretenaker},\ and\ \citenamefont
  {Goldfarb}}]{delpy2025anomalous}%
  \BibitemOpen
  \bibfield  {author} {\bibinfo {author} {\bibfnamefont {J.}~\bibnamefont
  {Delpy}}, \bibinfo {author} {\bibfnamefont {N.}~\bibnamefont {Fayard}},
  \bibinfo {author} {\bibfnamefont {F.}~\bibnamefont {Bretenaker}},\ and\
  \bibinfo {author} {\bibfnamefont {F.}~\bibnamefont {Goldfarb}},\ }\href@noop
  {} {\bibfield  {journal} {\bibinfo  {journal} {Physical Review Research}\
  }\textbf {\bibinfo {volume} {7}},\ \bibinfo {pages} {013298} (\bibinfo {year}
  {2025})}\BibitemShut {NoStop}%
\bibitem [{\citenamefont {Carl{\'e}}\ \emph {et~al.}(2021)\citenamefont
  {Carl{\'e}}, \citenamefont {Petersen}, \citenamefont {Passilly},
  \citenamefont {Hafiz}, \citenamefont {de~Clercq},\ and\ \citenamefont
  {Boudot}}]{carle2021exploring}%
  \BibitemOpen
  \bibfield  {author} {\bibinfo {author} {\bibfnamefont {C.}~\bibnamefont
  {Carl{\'e}}}, \bibinfo {author} {\bibfnamefont {M.}~\bibnamefont {Petersen}},
  \bibinfo {author} {\bibfnamefont {N.}~\bibnamefont {Passilly}}, \bibinfo
  {author} {\bibfnamefont {M.~A.}\ \bibnamefont {Hafiz}}, \bibinfo {author}
  {\bibfnamefont {E.}~\bibnamefont {de~Clercq}},\ and\ \bibinfo {author}
  {\bibfnamefont {R.}~\bibnamefont {Boudot}},\ }\href@noop {} {\bibfield
  {journal} {\bibinfo  {journal} {IEEE Transactions on Ultrasonics,
  Ferroelectrics, and Frequency Control}\ }\textbf {\bibinfo {volume} {68}},\
  \bibinfo {pages} {3249} (\bibinfo {year} {2021})}\BibitemShut {NoStop}%
\bibitem [{\citenamefont {Liu}\ \emph {et~al.}(2013)\citenamefont {Liu},
  \citenamefont {Merolla}, \citenamefont {Gu{\'e}randel}, \citenamefont
  {De~Clercq},\ and\ \citenamefont {Boudot}}]{liu2013ramsey}%
  \BibitemOpen
  \bibfield  {author} {\bibinfo {author} {\bibfnamefont {X.}~\bibnamefont
  {Liu}}, \bibinfo {author} {\bibfnamefont {J.-M.}\ \bibnamefont {Merolla}},
  \bibinfo {author} {\bibfnamefont {S.}~\bibnamefont {Gu{\'e}randel}}, \bibinfo
  {author} {\bibfnamefont {E.}~\bibnamefont {De~Clercq}},\ and\ \bibinfo
  {author} {\bibfnamefont {R.}~\bibnamefont {Boudot}},\ }\href@noop {}
  {\bibfield  {journal} {\bibinfo  {journal} {Optics express}\ }\textbf
  {\bibinfo {volume} {21}},\ \bibinfo {pages} {12451} (\bibinfo {year}
  {2013})}\BibitemShut {NoStop}%
\bibitem [{\citenamefont {Orszag}(2024)}]{orszag2024quantum}%
  \BibitemOpen
  \bibfield  {author} {\bibinfo {author} {\bibfnamefont {M.}~\bibnamefont
  {Orszag}},\ }\href@noop {} {\emph {\bibinfo {title} {Quantum optics:
  including noise reduction, trapped ions, quantum trajectories, and
  decoherence}}}\ (\bibinfo  {publisher} {Springer Nature},\ \bibinfo {year}
  {2024})\BibitemShut {NoStop}%
\bibitem [{\citenamefont {Khan}\ \emph {et~al.}(2022)\citenamefont {Khan},
  \citenamefont {Agarwalla},\ and\ \citenamefont {Jain}}]{khan2022quantum}%
  \BibitemOpen
  \bibfield  {author} {\bibinfo {author} {\bibfnamefont {S.}~\bibnamefont
  {Khan}}, \bibinfo {author} {\bibfnamefont {B.~K.}\ \bibnamefont
  {Agarwalla}},\ and\ \bibinfo {author} {\bibfnamefont {S.}~\bibnamefont
  {Jain}},\ }\href@noop {} {\bibfield  {journal} {\bibinfo  {journal} {Physical
  Review A}\ }\textbf {\bibinfo {volume} {106}},\ \bibinfo {pages} {022214}
  (\bibinfo {year} {2022})}\BibitemShut {NoStop}%
\bibitem [{sup()}]{supp}%
  \BibitemOpen
  \href@noop {} {}\bibinfo {note} {See Supplemental Material at
  https://journals.aps.org/authors/xxxx for the derivation of the functionsa a
  detailed description of the experimental setup.}\BibitemShut {Stop}%
\bibitem [{\citenamefont {Talker}\ \emph {et~al.}(2020)\citenamefont {Talker},
  \citenamefont {Zektzer}, \citenamefont {Barash}, \citenamefont {Mazurski},\
  and\ \citenamefont {Levy}}]{talker2020atomic}%
  \BibitemOpen
  \bibfield  {author} {\bibinfo {author} {\bibfnamefont {E.}~\bibnamefont
  {Talker}}, \bibinfo {author} {\bibfnamefont {R.}~\bibnamefont {Zektzer}},
  \bibinfo {author} {\bibfnamefont {Y.}~\bibnamefont {Barash}}, \bibinfo
  {author} {\bibfnamefont {N.}~\bibnamefont {Mazurski}},\ and\ \bibinfo
  {author} {\bibfnamefont {U.}~\bibnamefont {Levy}},\ }\href@noop {} {\bibfield
   {journal} {\bibinfo  {journal} {Journal of Vacuum Science \& Technology B}\
  }\textbf {\bibinfo {volume} {38}} (\bibinfo {year} {2020})}\BibitemShut
  {NoStop}%
\bibitem [{\citenamefont {of~Colorado~U.S.A.}(2025)}]{PrecisionGlassblowing}%
  \BibitemOpen
  \bibfield  {author} {\bibinfo {author} {\bibfnamefont {P.~G.}\ \bibnamefont
  {of~Colorado~U.S.A.}},\ }\href
  {https://precisionglassblowing.com/gas-wavelength-reference-cells/} {\bibinfo
  {title} {Gas wavelength reference cells}} (\bibinfo {year}
  {2025})\BibitemShut {NoStop}%
\bibitem [{\citenamefont {Orr}\ \emph {et~al.}(2018)\citenamefont {Orr},
  \citenamefont {Ishai},\ and\ \citenamefont {Roth}}]{orr2018high}%
  \BibitemOpen
  \bibfield  {author} {\bibinfo {author} {\bibfnamefont {G.}~\bibnamefont
  {Orr}}, \bibinfo {author} {\bibfnamefont {P.~B.}\ \bibnamefont {Ishai}},\
  and\ \bibinfo {author} {\bibfnamefont {M.}~\bibnamefont {Roth}},\ }\href@noop
  {} {\bibfield  {journal} {\bibinfo  {journal} {Measurement Science and
  Technology}\ }\textbf {\bibinfo {volume} {29}},\ \bibinfo {pages} {105502}
  (\bibinfo {year} {2018})}\BibitemShut {NoStop}%
\bibitem [{\citenamefont {Orr}\ and\ \citenamefont {Roth}(2012)}]{orr2012safe}%
  \BibitemOpen
  \bibfield  {author} {\bibinfo {author} {\bibfnamefont {G.}~\bibnamefont
  {Orr}}\ and\ \bibinfo {author} {\bibfnamefont {M.}~\bibnamefont {Roth}},\
  }\href@noop {} {\bibfield  {journal} {\bibinfo  {journal} {Review of
  scientific instruments}\ }\textbf {\bibinfo {volume} {83}} (\bibinfo {year}
  {2012})}\BibitemShut {NoStop}%
\bibitem [{\citenamefont {Debs}\ \emph {et~al.}(2008)\citenamefont {Debs},
  \citenamefont {Robins}, \citenamefont {Lance}, \citenamefont {Kruger},\ and\
  \citenamefont {Close}}]{debs2008piezo}%
  \BibitemOpen
  \bibfield  {author} {\bibinfo {author} {\bibfnamefont {J.}~\bibnamefont
  {Debs}}, \bibinfo {author} {\bibfnamefont {N.}~\bibnamefont {Robins}},
  \bibinfo {author} {\bibfnamefont {A.}~\bibnamefont {Lance}}, \bibinfo
  {author} {\bibfnamefont {M.}~\bibnamefont {Kruger}},\ and\ \bibinfo {author}
  {\bibfnamefont {J.}~\bibnamefont {Close}},\ }\href@noop {} {\bibfield
  {journal} {\bibinfo  {journal} {Applied optics}\ }\textbf {\bibinfo {volume}
  {47}},\ \bibinfo {pages} {5163} (\bibinfo {year} {2008})}\BibitemShut
  {NoStop}%
\bibitem [{\citenamefont {Wan}\ \emph {et~al.}(2016)\citenamefont {Wan},
  \citenamefont {Liu},\ and\ \citenamefont {Wang}}]{wan2016laser}%
  \BibitemOpen
  \bibfield  {author} {\bibinfo {author} {\bibfnamefont {J.-H.}\ \bibnamefont
  {Wan}}, \bibinfo {author} {\bibfnamefont {C.}~\bibnamefont {Liu}},\ and\
  \bibinfo {author} {\bibfnamefont {Y.-H.}\ \bibnamefont {Wang}},\ }\href@noop
  {} {\bibfield  {journal} {\bibinfo  {journal} {Chinese Physics B}\ }\textbf
  {\bibinfo {volume} {25}},\ \bibinfo {pages} {044204} (\bibinfo {year}
  {2016})}\BibitemShut {NoStop}%
\bibitem [{\citenamefont {Dammalapati}\ \emph {et~al.}(2009)\citenamefont
  {Dammalapati}, \citenamefont {Norris},\ and\ \citenamefont
  {Riis}}]{dammalapati2009saturated}%
  \BibitemOpen
  \bibfield  {author} {\bibinfo {author} {\bibfnamefont {U.}~\bibnamefont
  {Dammalapati}}, \bibinfo {author} {\bibfnamefont {I.}~\bibnamefont
  {Norris}},\ and\ \bibinfo {author} {\bibfnamefont {E.}~\bibnamefont {Riis}},\
  }\href@noop {} {\bibfield  {journal} {\bibinfo  {journal} {Journal of Physics
  B: Atomic, Molecular and Optical Physics}\ }\textbf {\bibinfo {volume}
  {42}},\ \bibinfo {pages} {165001} (\bibinfo {year} {2009})}\BibitemShut
  {NoStop}%
\bibitem [{\citenamefont {Liang}\ \emph {et~al.}(2018)\citenamefont {Liang},
  \citenamefont {Xu},\ and\ \citenamefont {Lin}}]{liang2018laser}%
  \BibitemOpen
  \bibfield  {author} {\bibinfo {author} {\bibfnamefont {S.-q.}\ \bibnamefont
  {Liang}}, \bibinfo {author} {\bibfnamefont {Y.-f.}\ \bibnamefont {Xu}},\ and\
  \bibinfo {author} {\bibfnamefont {Q.}~\bibnamefont {Lin}},\ }\href@noop {}
  {\bibfield  {journal} {\bibinfo  {journal} {Journal of Zhejiang
  University-SCIENCE A}\ }\textbf {\bibinfo {volume} {19}},\ \bibinfo {pages}
  {171} (\bibinfo {year} {2018})}\BibitemShut {NoStop}%
\bibitem [{\citenamefont {Nakayama}\ \emph {et~al.}(2010)\citenamefont
  {Nakayama}, \citenamefont {Yoshikawa}, \citenamefont {Matsumoto},
  \citenamefont {Torii},\ and\ \citenamefont {Kuga}}]{nakayama2010precise}%
  \BibitemOpen
  \bibfield  {author} {\bibinfo {author} {\bibfnamefont {K.}~\bibnamefont
  {Nakayama}}, \bibinfo {author} {\bibfnamefont {Y.}~\bibnamefont {Yoshikawa}},
  \bibinfo {author} {\bibfnamefont {H.}~\bibnamefont {Matsumoto}}, \bibinfo
  {author} {\bibfnamefont {Y.}~\bibnamefont {Torii}},\ and\ \bibinfo {author}
  {\bibfnamefont {T.}~\bibnamefont {Kuga}},\ }\href@noop {} {\bibfield
  {journal} {\bibinfo  {journal} {Optics express}\ }\textbf {\bibinfo {volume}
  {18}},\ \bibinfo {pages} {6604} (\bibinfo {year} {2010})}\BibitemShut
  {NoStop}%
\bibitem [{\citenamefont {Lente}(2025)}]{lente2025direction}%
  \BibitemOpen
  \bibfield  {author} {\bibinfo {author} {\bibfnamefont {G.}~\bibnamefont
  {Lente}},\ }\href@noop {} {\bibfield  {journal} {\bibinfo  {journal} {Journal
  of Mathematical Chemistry}\ ,\ \bibinfo {pages} {1}} (\bibinfo {year}
  {2025})}\BibitemShut {NoStop}%
\bibitem [{\citenamefont {Breuer}\ and\ \citenamefont
  {Petruccione}(2002)}]{breuer2002theory}%
  \BibitemOpen
  \bibfield  {author} {\bibinfo {author} {\bibfnamefont {H.-P.}\ \bibnamefont
  {Breuer}}\ and\ \bibinfo {author} {\bibfnamefont {F.}~\bibnamefont
  {Petruccione}},\ }\href@noop {} {\emph {\bibinfo {title} {The theory of open
  quantum systems}}}\ (\bibinfo  {publisher} {OUP Oxford},\ \bibinfo {year}
  {2002})\BibitemShut {NoStop}%
\bibitem [{\citenamefont {Noh}\ and\ \citenamefont
  {Jhe}(2010)}]{noh2010analytic}%
  \BibitemOpen
  \bibfield  {author} {\bibinfo {author} {\bibfnamefont {H.-R.}\ \bibnamefont
  {Noh}}\ and\ \bibinfo {author} {\bibfnamefont {W.}~\bibnamefont {Jhe}},\
  }\href@noop {} {\bibfield  {journal} {\bibinfo  {journal} {Optics
  communications}\ }\textbf {\bibinfo {volume} {283}},\ \bibinfo {pages} {2353}
  (\bibinfo {year} {2010})}\BibitemShut {NoStop}%
\end{thebibliography}%

\newpage
\begin{widetext}
\begin{center}
\textbf{\large Supplemental Materials: Beyond Maxwell-Boltzmann statistics using confined vapor cells}
\end{center}
\end{widetext}
\setcounter{equation}{0}
\setcounter{figure}{0}
\setcounter{table}{0}
\setcounter{page}{1}
\makeatletter
\renewcommand{\theequation}{S\arabic{equation}}
\renewcommand{\thefigure}{S\arabic{figure}}

\part{\large{Different relaxation effects as a function of cell thickness}}

The interaction of the atoms with the light is described by the Hamiltonian
\begin{equation}
\frac{d\rho}{dt}=-\frac{i}{\hbar}\left[\mathcal{H},\rho\right]+\mathcal{L}[\rho]\label{eq:Hamiltonian}
\end{equation}
 Incorporating the Lindblad operator for the single emitter system
into the Hamiltonian \cite{breuer2002theory}
\begin{equation}
\mathcal{L}[\rho]=\underset{ij}{\sum}\left(C_{ij}\rho C_{ij}^{\dagger}-\frac{1}{2}\left(\rho C_{ij}^{\dagger}C_{ij}+C_{ij}^{\dagger}C_{ij}\rho\right)\right)\label{eq:Lindblad}
\end{equation}
This accounts for both the decay and decoherence of the emitter. The
operator has three nonzero terms $C_{ij}\equiv(\Gamma_{spont}+\Gamma_{dephase})|i\rangle\langle j|$,
where $|i\rangle$ and $|j\rangle$ represent energy states $i$ and
$j$ associated to $E_{i}$ and $E_{j}$ respectively, $\Gamma_{spont}+\Gamma_{dephase}$are
the spontaneous emission rates associated with the transition from
the excited states to the ground states. $\Gamma_{dephase}$ is a
non radiative decay due to wall collisions. We can express the $\Gamma_{dephase}$
as
\begin{equation}
\Gamma_{dephase}=\frac{\bar{v}}{\bar{l}}+\frac{v_{z}}{L/2}\label{eq:Dephasing}
\end{equation}
 Where $\bar{v}=\sqrt{\frac{2k_{B}T}{m_{Rb}}}$ is the most probable
velocity, $\bar{l}$ is the mean free path, $v_{z}$ is the atomic
velocity in the direction perpendicular to the cell walls, and $L$
is the cell thickness. The mean free path $\bar{l}=\frac{\bar{v}_{rel}}{\Gamma_{se}}$,
where $\bar{v}_{rel}$ is the relative thermal velocity of the rubidium
atoms, and the rate of spin exchange $\Gamma_{se}=\sqrt{2}\cdot\sigma_{se}\cdot n_{Rb}$
where $\sigma_{se}=2\times10^{-14}cm^{2}$  is the spin exchange
cross section, $\bar{v}=\left(\frac{4k_{B}T}{\pi\mu_{Rb}}\right)^{1/2}$
is the average relative velocity of the rubidum atoms ($\mu_{Rb}$
is the rubidium atom\textquoteright s mass working temperature of
$150\text{°C}$), and $n_{Rb}$ is the density of atoms in the cell.

\begin{figure}[H]
\begin{centering}
\includegraphics[scale=0.6]{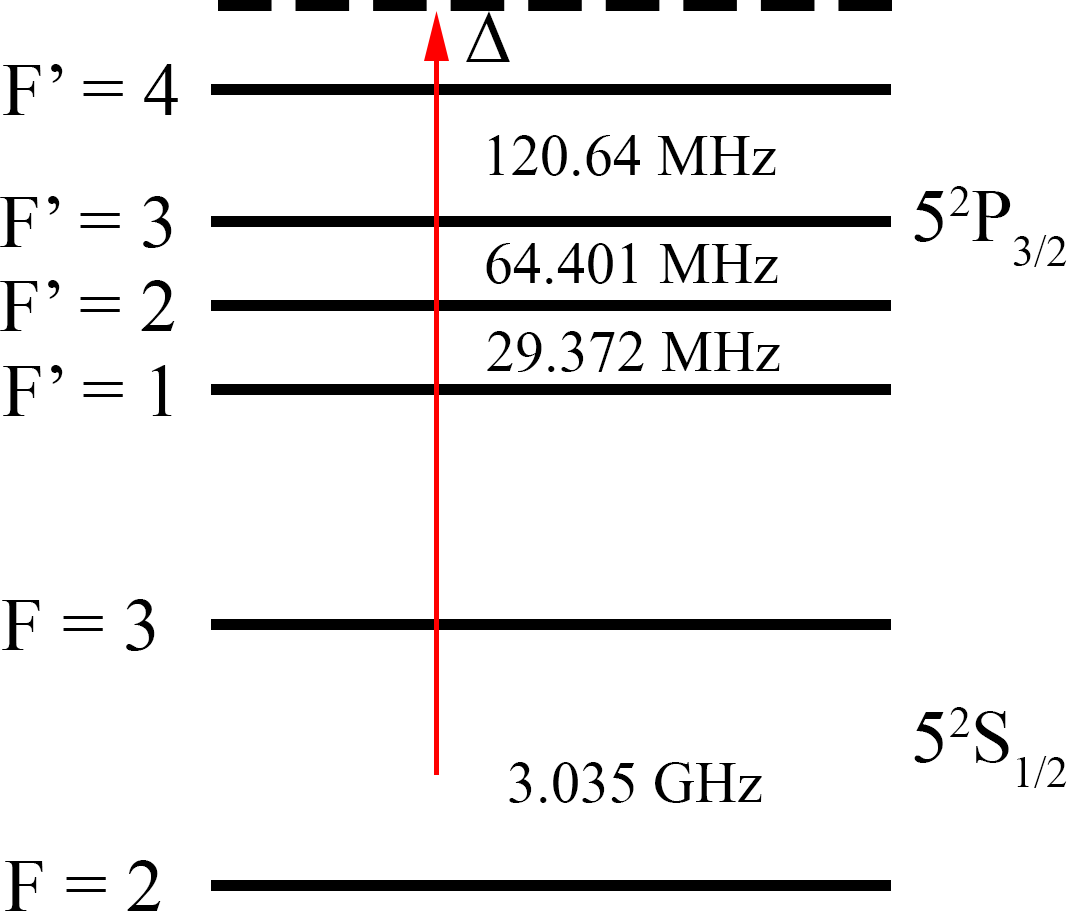}
\par\end{centering}
\caption{\label{fig:A-schematic-showing}A schematic showing the hyperfine
structure and intervals of $^{85}Rb$ for the $D_{2}$ spectroscopic
lines}

\end{figure}

To investigate the effect of relaxation processes as a function of
cell thickness, we measured the fluorescence spectra of the $D_{2}$
transition (Figure \ref{fig:A-schematic-showing}) from natural rubidium
vapor at a temperature of 150°C. We observed that reducing the cell thickness results in a noticeable decrease in the
spectral linewidth. This narrowing can be attributed to the increased
frequency of wall collisions, which preferentially filter out faster
atoms. Consequently, the signal is dominated by slower atoms, leading
to reduced Doppler broadening. A theoretical model solving Equations
\ref{eq:Hamiltonian}-\ref{eq:Dephasing} while taking the sum of
the excited population states and integrating them over the Doppler
broadening
\begin{equation}
Fluorescence=\stackrel[F=1]{4}{\sum}\int_{-\infty}^{\infty}\rho_{FF}(\Delta,v)W(v)dv
\end{equation}

where $W(v)$ is the one dimensional Maxwellian velocity distribution,
which is equal to
\begin{equation}
W(v)=\frac{1}{\sqrt{\pi u_{p}}}\exp\left(\frac{-v^{2}}{v_{p^{2}}}\right)
\end{equation}

$v_{p}=\sqrt{2k_{B}T/m}$ is the most probable velocity, $k_{B}$
is the Boltzmann's constant, $m$ is the mass of rubidium atoms and
$T$ is the temperature. 

\begin{figure*}
\begin{centering}
\includegraphics[scale=0.23]{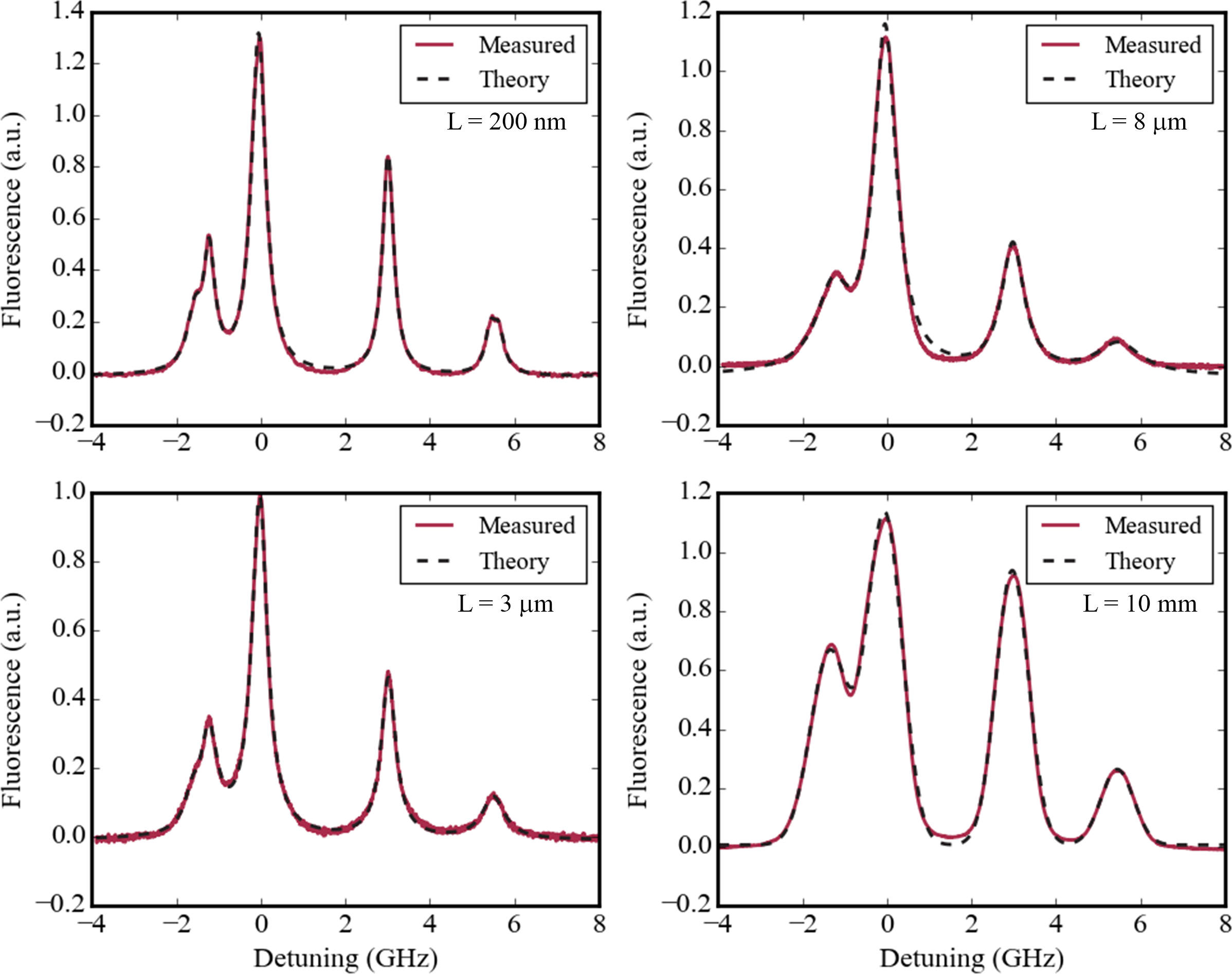}
\par\end{centering}
\caption{Comparison of fluorescence signal for the $D_{2}$ transition as a
function of frequency detuning the thin cell of four different values
of thickness ($T=150\text{°C}$).}

\end{figure*}

\part{\large{$g^{2}(\uptau)$derivation for submicron cells}}

The evolution of a damped two-level system, driven at resonances by
a coherent field $E_{0}cos(\omega t)$, can be described by the optical
Bloch equations \cite{noh2010analytic}
\begin{align}
\dot{\rho}_{22}(t) & =i\frac{\Omega}{2}\left[\rho_{12}(t)-\rho_{21}(t)\right]-\Gamma_{eff}\cdot\rho_{22}(t)\\
\dot{\rho}_{21}(t) & =i\frac{\Omega}{2}\left[\rho_{11}(t)-\rho_{22}(t)\right]-\left(\Gamma_{eff}-i\Delta)\rho_{21}(t)\right)\\
\dot{\rho1}_{21}(t) & =\left(\dot{\rho}_{21}(t)\right)^{\dagger}
\end{align}

\[
1=\dot{\rho}_{11}(t)+\dot{\rho}_{22}(t)
\]

Where $\Gamma_{eff}$ is define as
\begin{equation}
\Gamma_{eff}^{2}=\Gamma_{spont}^{2}+\left(\frac{\alpha}{t_{trans}}\right)^{2}\label{eq:effective_linewidth}
\end{equation}

Where $\Gamma_{nat}$ is the natural linewidth ( $2\pi\cdot6$ MHz)
and $\alpha$ is unitless factor that attenuates the transit time
broadening. This is much lower than the approximate value of $\alpha/t_{trans}\approx2\pi\cdot600$MHz.

The emission rate is given by
\begin{equation}
\Gamma_{2}=\Gamma_{eff}/2+\Gamma_{dephase}\label{eq:emission_rate}
\end{equation}
 and we defined the Rabi frequency to be $\Omega=d\cdot E/\hbar$.
$\rho_{ij}(t)$ are the density matrix elements, with $\rho_{11}$
being the population of the stable lower level, while the upper level
population $\rho_{22}$ has a decay rate of $\Gamma_{dephase}$. The
off-diagonal density matrix elements are damped at rate $\Gamma_{2}$,
and $\Delta$ is the detunning of the optical frequency from the resonant
frequency. The time dependence of the interaction has been eliminated
by making the rotating wave approximation and transforming to an appropriate
interaction picture. The steady state solution of these equations
gives the excited-state population as
\begin{equation}
\rho_{22}(\infty)=\frac{1}{2}\frac{\left(\frac{\Omega^{2}}{\Gamma_{eff}\cdot\Gamma_{2}}\right)}{1+\frac{\Delta^{2}}{\Gamma_{2}^{2}}+\left(\frac{\Omega^{2}}{\Gamma_{eff}\cdot\Gamma_{2}}\right)}
\end{equation}

Since $\Omega^{2}$ is proportional to the intensity $I$ of the light
that drives the excitation, we can express the relation between intensity
and Rabi frequency as 
\begin{equation}
\frac{\Omega^{2}}{\Gamma_{eff}\cdot\Gamma_{2}}=\frac{I}{I_{sat}}
\end{equation}

We do not turn to the time dependence of the density matrix which
we evaluate here for the special case of resonant excitation ($\Delta=0$).
By solving the coupled differential Equations 6-8
\begin{equation}
\rho_{22}(t)=a_{1}+a_{2}e^{-\frac{1}{2}(p-q)t}+a_{3}e^{-\frac{1}{2}(p+q)t}
\end{equation}

Where $p=\Gamma_{eff}+\Gamma_{2}$, $q=\sqrt{(\Gamma_{depth}-\Gamma_{2})^{2}-4\Omega^{2}}$

and the constants $a_{1},a_{2},a_{3}$ are
\begin{equation}
\left\{ a_{1},a_{2},a_{3}\right\} =\left\{ \frac{2\Omega^{2}}{p^{2}-q^{2}},-\frac{\Omega^{2}}{q(p-q)},\frac{\Omega^{2}}{q(P+q)}\right\} 
\end{equation}

The second order correlation function of the radiation field is defined
as \cite{orszag2024quantum}
\begin{equation}
g^{(2)}=\frac{\left\langle I(t)\cdot I(t+\uptau)\right\rangle }{\left\langle I(t)\right\rangle ^{2}}
\end{equation}

Where $I(t)$ is the intensity operator, and the whole function is
expressed in normal ordering. This is the normalized joint probability
of detecting a photon at time $t+\uptau$, given that another photon
was detected at time $t$. As shown in the main text the second order
correlation (Equation 1 in the main article) is equal to 
\begin{equation}
g^{(2)}(\uptau)=\frac{\rho_{22}(\uptau)}{\rho_{22}(\infty)}
\end{equation}

Hence the second order correlation at resonance 
\begin{equation}
g^{(2)}(\uptau)=1-\frac{p+q}{2q}e^{-\frac{1}{2}(p-q)\uptau}+\frac{p-q}{2q}e^{-\frac{1}{2}(p+q)\uptau}
\end{equation}

After simplifying the term and by taking the low light intensity limit
($\Omega\rightarrow0$) we can rewrite the second order correlation
as 
\begin{equation}
g^{(2)}(\uptau)=1+\frac{\Gamma_{eff}^{2}\Gamma_{2}e^{-\Gamma_{eff}\uptau}-\Gamma_{eff}(2\Gamma_{2}-\Gamma_{eff})^{2}e^{-\Gamma_{2}\uptau}}{(\Gamma_{eff}-\Gamma_{2})(2\Gamma_{2}+\Gamma_{eff})^{2}}\label{eq:second_order_correlation}
\end{equation}

\part{\large{Calculation of the coherence time (Numerical approach)}}

Let's compute the value at $\uptau=0$
\begin{equation}
g^{(2)}(0)=1+\frac{\Gamma_{eff}^{2}\Gamma_{2}-\Gamma_{eff}(2\Gamma_{2}-\Gamma_{eff})^{2}}{(\Gamma_{eff}-\Gamma_{2})(2\Gamma_{2}+\Gamma_{eff})^{2}}\label{eq:g^2(0)}
\end{equation}

Expanding the numerator 
\begin{widetext}
\begin{align}
(2\Gamma_{2}-\Gamma_{eff})^{2} & =4\Gamma_{2}^{2}-4\Gamma_{eff}\Gamma_{2}+\Gamma_{eff}^{2}\nonumber \\
\Rightarrow & \Gamma_{eff}^{2}\Gamma_{2}-\Gamma_{eff}(2\Gamma_{2}-\Gamma_{eff})^{2}=-4\Gamma_{eff}\Gamma_{2}^{2}+5\Gamma_{eff}^{2}\Gamma_{2}-\Gamma_{eff}^{3}\label{eq:numerator}
\end{align}
\end{widetext}

Hence, replacing the result in Equation \ref{eq:numerator} with the
numerator of Equation \ref{eq:g^2(0)} 
\begin{equation}
g^{(2)}(0)=1+\frac{-4\Gamma_{eff}\Gamma_{2}^{2}+5\Gamma_{eff}^{2}\Gamma_{2}-\Gamma_{eff}^{3}}{(\Gamma_{eff}-\Gamma_{2})(2\Gamma_{2}+\Gamma_{eff})^{2}}
\end{equation}

The peak height above the baseline is 
\begin{equation}
\Delta g^{(2)}=g^{(2)}(0)-1=\frac{5\Gamma_{eff}^{2}\Gamma_{2}-4\Gamma_{eff}\Gamma_{2}^{2}-\Gamma_{eff}^{3}}{(\Gamma_{eff}-\Gamma_{2})(2\Gamma_{2}+\Gamma_{eff})^{2}}
\end{equation}

Half the peak height above the baseline is 
\begin{equation}
\frac{\Delta g^{(2)}}{2}=\frac{1}{2}\frac{5\Gamma_{eff}^{2}\Gamma_{2}-4\Gamma_{eff}\Gamma_{2}^{2}-\Gamma_{eff}^{3}}{(\Gamma_{eff}-\Gamma_{2})(2\Gamma_{2}+\Gamma_{eff})^{2}}
\end{equation}

The value of $\frac{1}{2}g^{(2)}(0)$ at half maximum is 
\begin{equation}
\frac{1}{2}g^{(2)}(0)=1+\frac{\Delta g^{(2)}}{2}=1+\frac{1}{2}\frac{5\Gamma_{eff}^{2}\Gamma_{2}-4\Gamma_{eff}\Gamma_{2}^{2}-\Gamma_{eff}^{3}}{(\Gamma_{eff}-\Gamma_{2})(2\Gamma_{2}+\Gamma_{eff})^{2}}\label{eq:value of g(0) at half maximum}
\end{equation}

In order to find the width at FWHM we find the $\uptau>0$ for which
$g^{(2)}(\uptau)$ is equal to the expression of the height at half
the maximum i.e. $\frac{1}{2}g^{(2)}(0)$ (Equation \ref{eq:value of g(0) at half maximum})
\begin{equation}
g^{(2)}(\uptau)=1+\frac{1}{2}\frac{5\Gamma_{eff}^{2}\Gamma_{2}-4\Gamma_{eff}\Gamma_{2}^{2}-\Gamma_{eff}^{3}}{(\Gamma_{eff}-\Gamma_{2})(2\Gamma_{2}+\Gamma_{eff})^{2}}
\end{equation}

Substituting $g^{(2)}(\uptau)$ with it's expression (Equation \ref{eq:second_order_correlation})
and removing the baseline
\begin{widetext}
\begin{equation}
\frac{\Gamma_{eff}^{2}\Gamma_{2}e^{-\Gamma_{eff}\uptau}-\Gamma_{eff}(2\Gamma_{2}-\Gamma_{eff})^{2}e^{-\Gamma_{2}\uptau}}{(\Gamma_{eff}-\Gamma_{2})(2\Gamma_{2}+\Gamma_{eff})^{2}}=\frac{1}{2}\frac{5\Gamma_{eff}^{2}\Gamma_{2}-4\Gamma_{eff}\Gamma_{2}^{2}-\Gamma_{eff}^{3}}{(\Gamma_{eff}-\Gamma_{2})(2\Gamma_{2}+\Gamma_{eff})^{2}}\label{eq:comparing_the_correlation_function_to_half_peak_height}
\end{equation}
\end{widetext}

\subsection{First case: $L\ll\bar{l}$, $\Gamma_{2}\gg\Gamma_{eff}$ }

The analytical solution of Equation \ref{eq:comparing_the_correlation_function_to_half_peak_height}
is relatively complex, but by comparing the cell thickness to the
mean free path we can simplify the solution. In order to simplify
the calculation lets consider a case in which the quenching rate from
atom wall collisions $\Gamma_{dephase}$ (Equation \ref{eq:Dephasing})
is the dominant factor of the emission rate $\Gamma_{2}$ (Equation
\ref{eq:emission_rate}) hence for $\Gamma_{2}\gg\Gamma_{eff}$, requires
that $\frac{\bar{v}}{\bar{l}}\ll\frac{v_{z}}{L/2}$ meaning that the
mean free path is much larger than the cell thickness. This assumption
becomes valid as our cell thickness decreases into the micron and
sub micron. Based on the above assumption we can extract the value
of $\uptau$ at half the peak height, from Equation \ref{eq:comparing_the_correlation_function_to_half_peak_height}.
Reducing the denominator taking into account that $\Gamma_{2}\gg\Gamma_{eff}$
\[
(\bcancel{\Gamma_{eff}}-\Gamma_{2})(2\Gamma_{2}+\bcancel{\Gamma_{eff}})^{2}\approx-4\Gamma_{2}^{3}
\]
Let's examine the left term of Equation \ref{eq:comparing_the_correlation_function_to_half_peak_height}
\begin{widetext}
\begin{align}
\frac{\Gamma_{eff}^{2}\Gamma_{2}e^{-\Gamma_{eff}\uptau}-\Gamma_{eff}(2\Gamma_{2}-\Gamma_{eff})^{2}e^{-\Gamma_{2}\uptau}}{(\Gamma_{eff}-\Gamma_{2})(2\Gamma_{2}+\Gamma_{eff})^{2}} & \approx\frac{\Gamma_{eff}^{2}\Gamma_{2}e^{-\Gamma_{eff}\uptau}-\Gamma_{eff}(2\Gamma_{2}-\bcancel{\Gamma_{eff}})^{2}e^{-\Gamma_{2}\uptau}}{-4\Gamma_{2}^{3}}\approx\label{eq:partially_reduced_correlation_function}\\
\approx\bcancel{-\frac{1}{4}\frac{\Gamma_{eff}^{2}}{\Gamma_{2}^{2}}e^{-\Gamma_{eff}\uptau}} & +\frac{\Gamma_{eff}}{\Gamma_{2}}e^{-\Gamma_{2}\uptau}
\end{align}
\end{widetext}

where the last term $\frac{\Gamma_{eff}}{\Gamma_{2}}e^{-\Gamma_{2}\uptau}$
is positive, significantly more dominant, and drops faster than $\frac{\Gamma_{eff}^{2}}{\Gamma_{2}^{2}}e^{-\Gamma_{eff}\uptau}$
. Hence, replacing this result with the correlation factor in Equation
\ref{eq:second_order_correlation} we obtain the reduced second order
correlation in ultrathin cells
\begin{equation}
g^{(2)}(\uptau)\approx1+\frac{\Gamma_{eff}}{\Gamma_{2}}e^{-\Gamma_{2}\uptau}\label{eq:reduced_second_order_correlation}
\end{equation}

Returning to the right term of Equation \ref{eq:comparing_the_correlation_function_to_half_peak_height}
\begin{widetext}
\begin{align}
\frac{1}{2}\frac{5\Gamma_{eff}^{2}\Gamma_{2}-4\Gamma_{eff}\Gamma_{2}^{2}-\Gamma_{eff}^{3}}{(\Gamma_{eff}-\Gamma_{2})(2\Gamma_{2}+\Gamma_{eff})^{2}} & \approx\frac{1}{8}\frac{4\Gamma_{eff}\Gamma_{2}^{2}+\Gamma_{eff}^{3}-5\Gamma_{eff}^{2}\Gamma_{2}}{\Gamma_{2}^{3}} = \frac{1}{8}\left(\frac{4\Gamma_{eff}}{\Gamma_{2}}+\bcancel{\frac{\Gamma_{eff}^{3}}{\Gamma_{2}^{3}}}-\bcancel{\frac{5\Gamma_{eff}^{2}}{\Gamma_{2}^{2}}}\right)\approx\label{eq:delta_of_reduced_half_maximum}\\
\approx & \frac{1}{2}\frac{\Gamma_{eff}}{\Gamma_{2}}\nonumber 
\end{align}
\end{widetext}
we can reduce the higher terms as $1\gg\Gamma_{eff}/\Gamma_{2}\Rightarrow\Gamma_{eff}/\Gamma_{2}\ggg\Gamma_{eff}^{2}/\Gamma_{2}^{2}\ggg\Gamma_{eff}^{3}/\Gamma_{2}^{3}$.
The time delay $\uptau_{\nicefrac{1}{2}}$ at half the height of $g^{(2)}(0)$
allows us to evaluate the spread that we can calculate it using the
results of Equations \ref{eq:partially_reduced_correlation_function}
and \ref{eq:delta_of_reduced_half_maximum}
\[
\frac{\Gamma_{eff}}{\Gamma_{2}}e^{-\Gamma_{2}\uptau}=\frac{1}{2}\frac{\Gamma_{eff}}{\Gamma_{2}}
\]
\begin{align*}
e^{-\Gamma_{2}\uptau} & =2^{-1}\\
\Gamma_{2}\uptau= & \ln2\\
\Rightarrow\uptau= & \frac{\ln2}{\Gamma_{2}}
\end{align*}
As the linewidth is twice $\uptau_{\nicefrac{1}{2}}$
\begin{equation}
FWHM=\frac{2\ln2}{\Gamma_{2}}\propto\frac{1.386}{v_{z}}\cdot L\label{eq:FWHM}
\end{equation}

\subsection{Second case: $\Gamma_{eff}\gg\Gamma_{dephase}$}

When dealing with large cell thicknesses, non radiative decay due
to wall collisions is negligible compared to the effective decay hence
the emission rate (Equation \ref{eq:emission_rate}) is approximately
\[
\Gamma_{2}\approx\frac{\Gamma_{eff}}{2}
\]

the denominator of Equation \ref{eq:comparing_the_correlation_function_to_half_peak_height}
becomes
\[
(\Gamma_{eff}-\Gamma_{2})(2\Gamma_{2}+\Gamma_{eff})^{2}\approx\frac{1}{2}\Gamma_{eff}(2\Gamma_{eff})^{2}=2\Gamma_{eff}^{3}
\]

the right side of Equation \ref{eq:comparing_the_correlation_function_to_half_peak_height}
becomes 
\[
\frac{1}{2}\frac{\frac{5}{2}\bcancel{\Gamma_{eff}^{3}}-2\bcancel{\Gamma_{eff}^{3}}-\bcancel{\Gamma_{eff}^{3}}}{2\bcancel{\Gamma_{eff}^{3}}}=\frac{5}{8}
\]

and the left side of Equation \ref{eq:comparing_the_correlation_function_to_half_peak_height}
\begin{widetext}
\begin{align}
\approx\frac{1}{2}\frac{\bcancel{\Gamma_{eff}^{2}}\cdot\frac{1}{2}\bcancel{\Gamma_{eff}}\cdot e^{\Gamma_{eff}\uptau}-\Gamma_{eff}\bcancel{(\Gamma_{eff}-\Gamma_{eff})^{2}e^{-\frac{1}{2}\Gamma_{eff}\uptau}}}{2\bcancel{\Gamma_{eff}^{3}}}=\frac{1}{4}e^{-\Gamma_{eff}\uptau}\\
\frac{1}{4}e^{-\Gamma_{eff}\uptau} & =\frac{5}{8}\\
\Gamma_{eff}\uptau & =\ln5-\ln2=0.9163
\end{align}
\end{widetext}
given that $\Gamma_{eff}\approx2\pi\times\frac{\alpha\bar{v}}{D}=2\pi\times200\times10^{6}Hz$
for values of $\alpha=8.25,\,\bar{v}=280\,m/s,\,D=7\mu m$.
\begin{equation}
\uptau\approx\frac{0.9163}{\Gamma_{eff}}=4.419\times10^{-10}\,s
\end{equation}

\part{\large{Detailed description of the experimental setup}}

The full experimental setup, illustrated in Fig. S4, was used throughout
the experiment. A Toptica DL pro laser operating at 780\,nm with an
output power of 80\,mW provided the required spatial and temporal
stability for the beam entering the thin vapor cell. To suppress beam
jitter and eliminate higher-order spatial modes, the laser beam was
passed through a high-finesse, three-mirror ring cavity acting as
a mode cleaner. The incident beam typically includes significant contributions
from higher-order modes---particularly first- and second order which
distort the beam profile. The length of the cavity was adjusted to
ensure resonance only for the fundamental TEM\textsubscript{0}\textsubscript{0}
mode. 

\begin{figure*}
\begin{centering}
\includegraphics[scale=0.23]{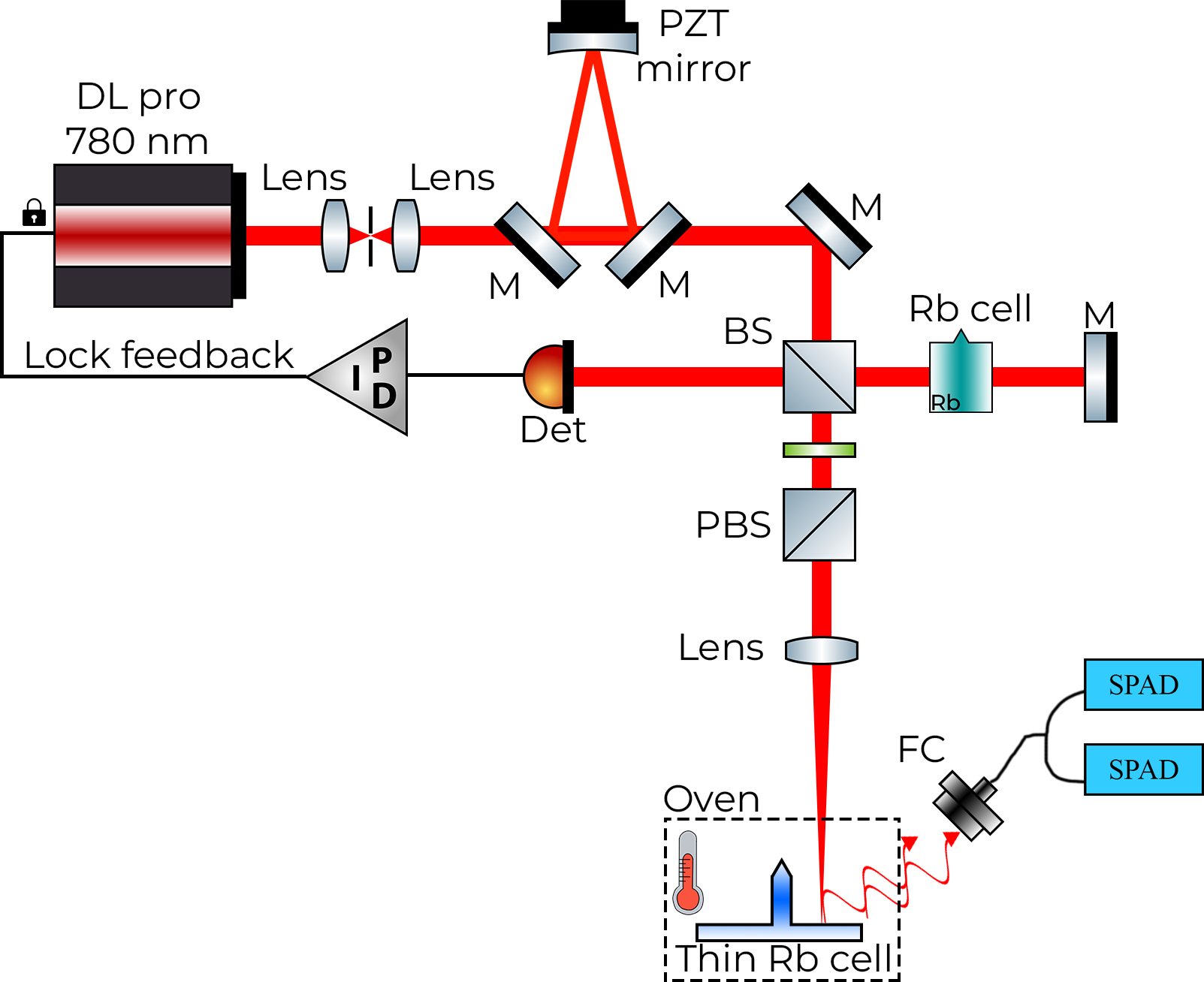}
\par\end{centering}
\caption{\label{fig:Experimental-setup-used}Experimental setup used for the
intensity auto-correlation measurements for our thin cell. M: mirror;
PBS: polarizing beam splitter; FC: fiber collimator; Det: detector.}

\end{figure*}

In addition to the mode-cleaner cavity, a spatial filter composed
of two lenses and an iris was used to further improve the beam quality.
The laser beam was then split into two paths: one arm was directed
to a saturated absorption spectroscopy setup for frequency stabilization,
while the second arm was directed toward the experiment. This beam
passed through a half-wave plate and a polarizing beam splitter to
control its intensity and ensure clean linear polarization. After
focusing the beam into the vapor cell using a lens, the fluorescence
signal was collected with a fiber collimator. The collected light
was then split via a fiber splitter and directed to two single-photon
detectors, enabling measurement of the fluorescence intensity auto-correlation
function.


\end{document}